\documentclass[%
letter,
groupedaddress,
showpacs,
 amsmath,amssymb,
 aps,
pre,
]{revtex4-1}

\usepackage{graphicx}
\usepackage{dcolumn}
\usepackage{bm}
\usepackage{epstopdf}
\usepackage{colortbl}
\usepackage{float}

\newcommand{\vel}{\bm{{u}}}

\begin{document}
\title{Weakly  {\color{black} non-Boussinesq} convection in a gaseous spherical shell}

\author{Lydia Korre}
\author{Nicholas Brummell}
\author{Pascale Garaud}
\affiliation{Department of Applied Mathematics and Statistics, Jack Baskin School of Engineering, University of California Santa Cruz, 1156 High Street, Santa Cruz, CA
95064, USA}

\date{\today}

\begin{abstract}
We examine the dynamics associated with  weakly compressible convection in a spherical shell by running 3D {\color{black}direct numerical simulations}  using the Boussinesq formalism \cite{SV}. Motivated by problems in astrophysics, we assume the existence of a  finite adiabatic temperature gradient $\nabla T_{\rm{ad}}$ and use mixed boundary conditions for the temperature with  fixed flux at the inner boundary and fixed temperature at the outer boundary. This  setup is intrinsically  more asymmetric than the more standard case of Rayleigh-B\'{e}nard convection in liquids between parallel plates with fixed temperature boundary conditions. {\color{black} Conditions where there is substantial asymmetry can cause  a dramatic change in the nature of convection and we demonstrate that this is the case here}. The flows can become {\color{black}pressure-} rather than {\color{black}buoyancy-} dominated leading to anomalous heat transport by upflows. Counter-intuitively, the background temperature gradient $\nabla\bar{T}$ can develop a  subadiabatic layer (where $\bm{g}\cdot\nabla\bar{T}<\bm{g}\cdot\nabla T_{\rm{ad}}$, where $\bm{g}$ is  gravity) although convection remains vigorous at every point across the shell. This indicates a high degree of non-locality.
\end{abstract}


\maketitle

\section{INTRODUCTION}
\par Convection is a ubiquitous physical process in  geophysical fluid dynamics, which has been extensively studied analytically, experimentally and numerically because of the vital role it plays in the global dynamics  of the Earth's mantle (e.g. \cite{BSG89,GS93,exp} and for a review see \cite{mantle} and references therein), oceans \cite{ocean} and  atmosphere \cite{atmo}.
Convection is also important in astrophysical settings such as the gaseous interiors of stars and planets where the convective zones are usually global, either spanning the entire object or at least a deep spherical shell. By contrast with geophysical convection, relatively little is known about convection in astrophysical objects. Observationally speaking, a limited amount of information can be obtained either through  direct imaging of the surface (e.g. see  \cite{RR10} for a review), or indirectly using asteroseismology to infer, for instance, the mean temperature profile  within the convection zone \cite{astro}. Meaningful physical experiments are almost impossible to design  because the governing parameters appropriate to the interiors of stars and planets are vastly different from  those achievable in a laboratory. In particular, the Prandtl number, which is the ratio of the kinematic viscosity to the thermal diffusivity, is much smaller than unity in astrophysical plasmas (e.g.  $\sim \color{black}10^{-2}$ in giant planets, and $\sim 10^{-6}$ in stars)   whereas it is usually of order unity or much larger in geophysical applications. Among other things, this implies that the  ordering of the relevant dynamical timescales is different in the two regimes, and that the effects of the inertial terms in astrophysical convection are much larger than in geophysical convection. 

In this paper, we attempt to shed new light on the subject by investigating the dynamics of  convection in weakly compressible gaseous spherical shells in the low Prandtl {\color{black} number} parameter regime using {\color{black}direct numerical simulations} (DNSs) with mixed temperature boundary conditions ({\color{black}here meant to imply} fixed flux at the inner boundary and fixed temperature at the outer boundary). This setup is designed to capture some of the most salient features of convection in stars and giant planets, and differs in significant ways from most studies of convection to date.
\par Arguably, the most commonly studied form of convection is thermal Rayleigh-B\'{e}nard convection (RBC thereafter) between two parallel plates  where a Boussinesq liquid (in the original Boussinesq sense \cite{Bouss,Ober}) is heated from below and cooled from above, and the two rigid boundaries are held at constant temperatures. For sufficiently strong driving, as measured by the Rayleigh number, buoyancy forces  overcome thermal and viscous damping and turbulent heat transport by convection dominates conduction.  This highly symmetric idealized model setup has been studied extensively  in both 2D and 3D \cite{Veronis66, MW73,SS90,Julien, Celine} (also for a general review of RBC see \cite{review} and references therein).    

When studying geophysical  problems, several extensions of this basic model are usually considered depending on the specific application. Studies of mantle convection usually adopt a spherical shell geometry and consider the limit of infinite Prandtl number \cite{chandra,Busse,Zebib80,SZ80,Zebib85,Machetel,Bercovici,Vangelov,Jarvis,Choblet}. More generally, geophysically motivated studies of convection in spherical shells  sometimes include the effect of rotation or allow for a finite Prandtl number 
\cite{Dormy,Tilgner96,Feudel11,Gastine15, Gastine16} but have so far  nearly always used   fixed temperature boundary conditions.   The majority of these investigations have focused on the  derivation of scaling laws for global quantities such as the heat flux  or the total kinetic energy as functions of input parameters, as well as  developing models for the boundary layers {\color{black}(see \cite{GL2000} for a review)}.

\par In  astrophysical applications, on the other hand, the fluid is generally  compressible.  Solving the  compressible  Navier-Stokes equations requires the resolution of  timescales  associated with fast sound waves, as well as the much slower timescales associated with global thermal or viscous adjustment. This stiffness is a severe impediment to simulation and filtering out the fast sonic dynamics is often desirable.  
One  way of accounting for weak compressibility in astrophysical convection is through the anelastic approximation  \cite{Batchelor,Ogura,Gough69,Latour,GG}, which filters out sound waves while allowing for strong variations in the background density. This is the more commonly adopted formalism for the study of solar convection (\cite{Miesch} and references therein) and stellar convection (e.g. \cite{BBT,BBB,Aug12,BVZ})  but it has significant drawbacks.   First of all, there are numerous formulations of the approximation and there is some debate about their relative validity \cite{Vasil,ancom}.  Secondly, the anelastic approximation is usually based on the assumption of small departures from adiabaticity which is not guaranteed in all reasonable problems. 
\par Another commonly used approximation under which   sound waves are filtered out is  the Boussinesq approximation for gases \cite{SV}. It is important to note that the \textit{standard} Boussinesq approximation \cite{Bouss,Ober} should not be  used in astrophysical applications because of the compressibility of the gas (although it is still sometimes used for simplicity \cite{Gilman76,Gilman77, Gilman78,Christensen,Heimpel05,HA07,AH07}).  However, Spiegel \& Veronis (1960)  \cite{SV} (SV thereafter) showed that it is possible to generalize {\color{black}the approximation} to take into account some effects of compressibility, {\color{black}thereby} allowing its use in modeling convection in gaseous systems, such as the Earth's atmosphere or the interiors of stars and planets. Assuming that the size of the convective region is much smaller than any scale height of the system (including the local radius, if the convection zone is a spherical shell), and that the fluid motions are much slower than the local speed of sound, they showed that the only effect of compressibility is to heat or cool a parcel of fluid as it shrinks or expands to adjust to the ambient hydrostatic pressure. As a result, their formulation contains {\color{black} an additional} term in the temperature equation which is proportional to the  local adiabatic temperature gradient (which is non-zero for gases) but is otherwise identical to the traditional Boussinesq approximation. Studies of astrophysical convection in Cartesian geometry or in a very thin spherical shell  using the SV Boussinesq (SVB thereafter) approximation were presented for instance in  \cite{Cat03} and \cite{Miesch2001}.

\par In this work we propose to study 3D DNSs of low Prandtl number convection in a spherical geometry using the SVB equations {\color{black} with particular applications to solar and stellar convection in mind}.   We {\color{black}therefore} deviate from the usual assumption of fixed temperature boundary conditions and instead,  consider a more astrophysically realistic setup with fixed flux at the inner boundary and fixed temperature at the outer boundary.  Indeed, in stars like the Sun for instance, the flux through the base of an outer convection zone is fixed by the luminosity of the star, which in turn is set by the nuclear generation rate within the core. 
\par These four elements (spherical geometry, {\color{black} weak} compressibility, mixed temperature boundary conditions and low Prandtl number) have never, to the authors' knowledge, been used in conjunction and yet are all crucial elements of astrophysical convection. As we demonstrate in this work, their combined effect is to create substantial asymmetry between the upflows and downflows, which in turn transforms the nature of convection near the lower boundary, with surprising repercussions throughout the entire convection zone. 
\par  The paper is organized as follows: Sections II and III present two different model setups that both possess all four properties listed above, and report on the results of a suite of numerical simulations for varying Rayleigh number. Surprising new dynamics are observed, which are then analyzed and explained in detail in Section IV.   In Section V, we explore a third model setup which more closely resembles the Sun (although still simplistically), in order to test the robustness of our results and to assess the applicability of what we have found to {\color{black} the circumstances that most interest us}. Finally, in Section VI, we summarize our results and  discuss the possible limitations of the SVB approximation.

\section{Boussinesq convection in a weakly compressible spherical shell}
\subsection{Mathematical formulation}

We begin our systematic investigation of the effects of mixed temperature boundary conditions and weak compressibility on the dynamics of Rayleigh-B\'{e}nard convection in a spherical shell by constructing the simplest possible model with these properties. In this model, and in all of the ones that follow, we consider a spherical shell located between an inner sphere of radius $r_i$ and an outer sphere of radius $r_o$.  {\color{black} For simplicity,} we assume constant thermal expansion coefficient $\alpha$, viscosity $\nu$, thermal diffusivity $\kappa$, adiabatic temperature gradient $dT_{\rm{ad}}/dr$ and gravity $g$. In the absence of fluid motion and when the system is in a steady state, the background radiative temperature gradient is obtained by solving 
 \begin{equation}
 \label{rad}
 \kappa\nabla^2 T_{\rm{rad}}(r)=0\Rightarrow\kappa r^2\displaystyle\frac{dT_{\rm{rad}}}{dr}=\rm{const},
 \end{equation}
where $r$ is the local radius.
The inner fixed flux boundary condition implies that
\begin{equation}
-\kappa \frac{dT_{\rm rad}}{dr} \bigg|_{r=r_i}  = F_{\rm{rad}},  
\end{equation} 
where $F_{\rm{rad}}$ is the temperature flux per unit area through the inner boundary, whereas the outer fixed temperature boundary condition is  $T(r_o) = T_o$. Then, solving Eq. (\ref{rad})  using the first boundary condition implies that 
\begin{equation}
\displaystyle\frac{dT_{\rm{rad}}}{dr}=-\frac{F_{\rm{rad}}}{\kappa}\left(\frac{r_i}{r}\right)^2,
\end{equation}  
which, along with the second boundary condition, gives
\begin{equation}
\displaystyle T_{\rm{rad}}(r) =\frac{F_{\rm{rad}} r_i^2}{\kappa}\left(\frac{1}{r}-\frac{1}{r_o}\right)+T_o. 
\end{equation}
We clearly see that, {\color{black}in contrast to the Cartesian case}, the radiative temperature gradient in a spherical geometry is not constant but  depends on the radius. This implies in turn that $dT_{\rm{rad}}/dr - dT_{\rm{ad}}/dr$ also varies with depth. {\color{black} Note that for the SVB approximation to be valid, $\Delta T=T_{\rm{rad}}(r_i)-T_{\rm{rad}}(r_o)$ must be much smaller than, say, $T_o$. This is true either for small enough $r_o-r_i$ (thin layer) given $F_{\rm{rad}}$, or for small enough $F_{\rm{rad}}$ given $r_o-r_i$.}
\par We now let $T(r,\theta,\phi,t)=T_{\rm{rad}}(r)+\Theta(r,\theta,\phi,t)$ where $\Theta$ is the temperature perturbation to the radiative background.
We also  assume a linear relationship between the temperature and density perturbations consistent with the SVB approximation,  $\rho/\rho_m = - \alpha \Theta$, where $\rho_m$ is the mean density of the background fluid. With these assumptions, the governing SVB equations  are:  
\begin{equation}
\nabla\cdot \vel=0,
\end{equation}
\begin{equation}
\displaystyle\frac{\partial\vel}{\partial t}+\vel\cdot\nabla\vel=-\frac{1}{\rho_m}\nabla p+\alpha \Theta g\boldsymbol{e_r}+\nu\nabla^2\vel,
\end{equation}
and
\begin{equation}
\displaystyle\frac{\partial\Theta}{\partial t}+\vel\cdot\nabla\Theta+u_r\left(\frac{dT_{\rm{rad}}}{dr}-\frac{dT_{\rm{ad}}}{dr}\right)=\kappa\nabla^2\Theta{\color{black},}
\end{equation}
{\color{black} where $\vel = (u_r, u_{\theta}, u_{\phi})$ is the velocity field, and $p$ is the pressure.} We non-dimensionalize the problem by {\color{black} using \footnote{We numerically solve the non-dimensional Boussinesq equations in which we have used the outer radius  as the lengthscale. If we wanted to compare spherical numerical simulations with Cartesian ones, we would have to non-dimensionalize the problem using the thickness of the shell $[l]=r_o-r_i=L$ such that both the problems could have the same effective Rayleigh number for accurate comparison.}} $[l]=r_o$, $[t]=r_o^2/\nu$, $[u]=\nu/r_o$ and $[T]=|dT_o/dr-dT_{\rm{ad}}/dr| r_o$ as the unit length, time, velocity and temperature, where ${dT_o}/{dr}\equiv{dT_{\rm{rad}}}/{dr}|_{r=r_o}$. Then, we can write the non-dimensional equations as:
\begin{equation}
\label{divU}
{\nabla}\cdot{\vel}=0,
\end{equation}
\begin{equation}
\displaystyle\frac{\partial{\vel}}{\partial{t}}+{\vel}\cdot{\nabla}{\vel}=-{\nabla}{ p}+\frac{\text{Ra}_o}{\text{Pr}}{\Theta}\boldsymbol{e_r}+{\nabla}^2{\vel},
\end{equation}
and
\begin{equation}
\label{heq}
\displaystyle\frac{\partial{\Theta}}{\partial{ t}}+{\vel}\cdot{\nabla}{\Theta}+\beta({r}){u_r}=\frac{1}{\rm{Pr}}{{\nabla}^2{\Theta}}.
\end{equation}
All the variables and parameters are now implicitly non-dimensional, which introduces the Prandtl number Pr and the Rayleigh Ra$_o$ defined as
\begin{equation}
\text{Pr}=\displaystyle\frac{\nu}{\kappa}\quad \text{and}\quad \text{Ra}_o=\displaystyle\frac{\alpha g\left|\displaystyle\frac{dT_o}{dr}-\frac{dT_{\rm{ad}}}{dr}\right|r_o^4}{\kappa\nu},
\end{equation}
and  the non-dimensional superadiabaticity
\begin{equation}
\label{beta}
\beta(r)=\displaystyle\frac{\displaystyle\frac{dT_{\rm{rad}}}{dr}-\displaystyle\frac{dT_{\rm{ad}}}{dr}}{\displaystyle\left|\frac{dT_o}{dr}-\displaystyle\frac{dT_{\rm{ad}}}{dr}\right |}=\displaystyle\frac{\displaystyle\left(\frac{1}{r}\right)^2\frac{dT_{o}}{dr}-\displaystyle\frac{dT_{\rm{ad}}}{dr}}{\displaystyle\left|\frac{dT_o}{dr}-\displaystyle\frac{dT_{\rm{ad}}}{dr}\right|}.
\end{equation}
Another way to interpret $\beta$ is to note that it is minus the ratio of the local Rayleigh number Ra$(r)$ to Ra$_o$ i.e. 
\begin{equation}
\displaystyle\beta(r)=-\frac{\rm{Ra}(r)}{\rm{Ra}_o},
\end{equation}
where
 \begin{equation}
 \displaystyle \text{Ra}(r)=\frac{\alpha g\left|\displaystyle \frac{dT_{\rm{rad}}}{dr}-\frac{dT_{\rm{ad}}}{dr}\right|r_o^4}{\kappa\nu}{\color{black}=\frac{\alpha g\left|\displaystyle -\frac{F_{\rm{rad}}}{\kappa}\left(\frac{r_i}{r}\right)^2-\frac{dT_{\rm{ad}}}{dr}\right|r_o^4}{\kappa\nu}}.
 \end{equation}

Finally, note that while $\beta$ seems to depend on two dimensional quantities $dT_{\rm{rad}}/dr$ and $dT_{\rm{ad}}/dr$ (see Equation (\ref{beta})), it can be rewritten  in this simple model just in terms of a single non-dimensional parameter $\chi$, defined as
\begin{equation}
\label{chi}
\chi=\displaystyle\left|\frac{\displaystyle\frac{dT_o}{dr}-\displaystyle\frac{dT_{\rm{ad}}}{dr}}{\displaystyle\frac{dT_o}{dr}}\right|{\color{black}=\displaystyle\left|1+\dfrac{\kappa\dfrac{dT_{\rm{ad}}}{dr} r_o^2}{F_{\rm{rad}} r_i^2}\right|},
\end{equation}
so that
\begin{equation}
\label{eqbeta}
\beta(r)=\frac{1-\chi-(1/r)^2}{\chi}.
\end{equation}
Note that $\beta(1)=-1$ and $\beta(r_i/r_o)=(1-\chi-(r_o/r_i)^2)/\chi$.
\par
\begin{figure}[H]
\centering
\includegraphics[scale=0.3]{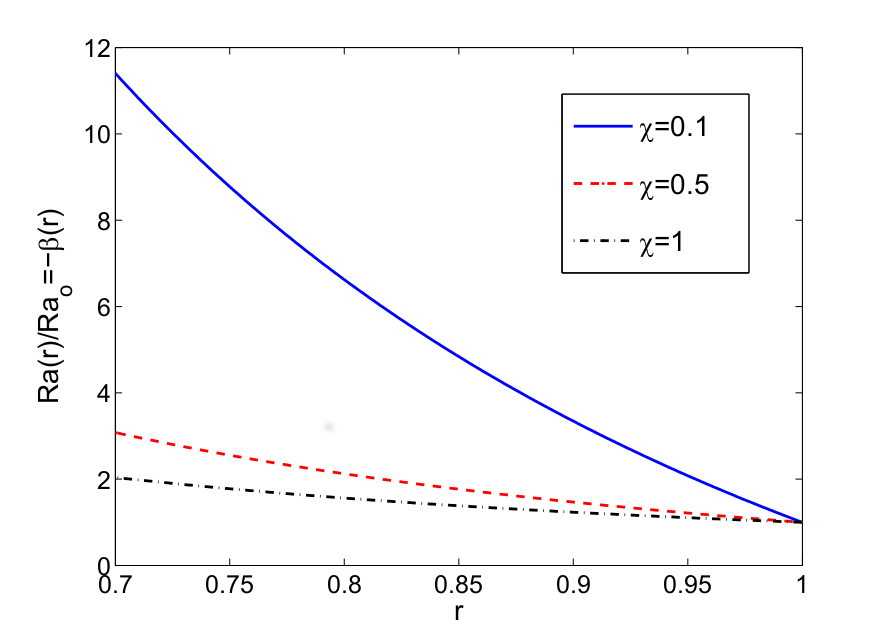}
\caption{\label{fig_1} (Color online) The dependence of {\color{black}$\beta(r)$}  on $\chi$.}
\end{figure}

  Figure \ref{fig_1} illustrates how $\beta$, and therefore the local Rayleigh ratio  Ra$(r)$/Ra$_o$, depends on $\chi$. 
Note that for $\chi=0.5$ the local Rayleigh number at the inner boundary is about 3 times larger than Ra$_o$, whereas for $\chi=0.1$, it is 11 times larger, illustrating that a small $\chi$ implies a stronger variation of the local Rayleigh number across the shell.
In the limit of a very thin shell $(r_i/r_o \rightarrow 1)$ on the other hand (which recovers the case of convection between infinite parallel plates), $\beta(r)$ {\color{black}tends to the constant $ -1$} regardless of $\chi$. The functional form of $\beta$ therefore depends on our choice of boundary conditions and on the fact that we are operating in an appreciably deep spherical shell (see Equation (\ref{eqbeta})).

\subsection{Numerical simulations}

\par In order to study the influence of weak compressibility and sphericity (which manifest themselves in a variable  $\beta(r)$), and of mixed temperature boundary conditions on the model dynamics, we have run 3D DNSs  solving   Equations (\ref{divU})-(\ref{heq}) in a spherical shell, exactly as outlined above, using the PARODY code \cite{parody}. The boundary conditions for the temperature perturbations  $\Theta$ are such that we have fixed flux  at the inner boundary,  $\partial\Theta/\partial r|_{r_i}=0$ and fixed temperature  at the outer boundary, $\Theta(r_o)=0$. The velocity boundary conditions are stress-free  at both the inner and outer  boundaries. The simulations discussed in this section are referred to as ``Model A" simulations. Table \ref{table1} summarizes our various runs in this setup, as well as those later discussed in Sections III and V. Note that $r_i/r_o=0.7$ and Pr $=0.1$ for all the simulations presented in this paper. \par
\begin{figure*}
\includegraphics[scale=0.25]{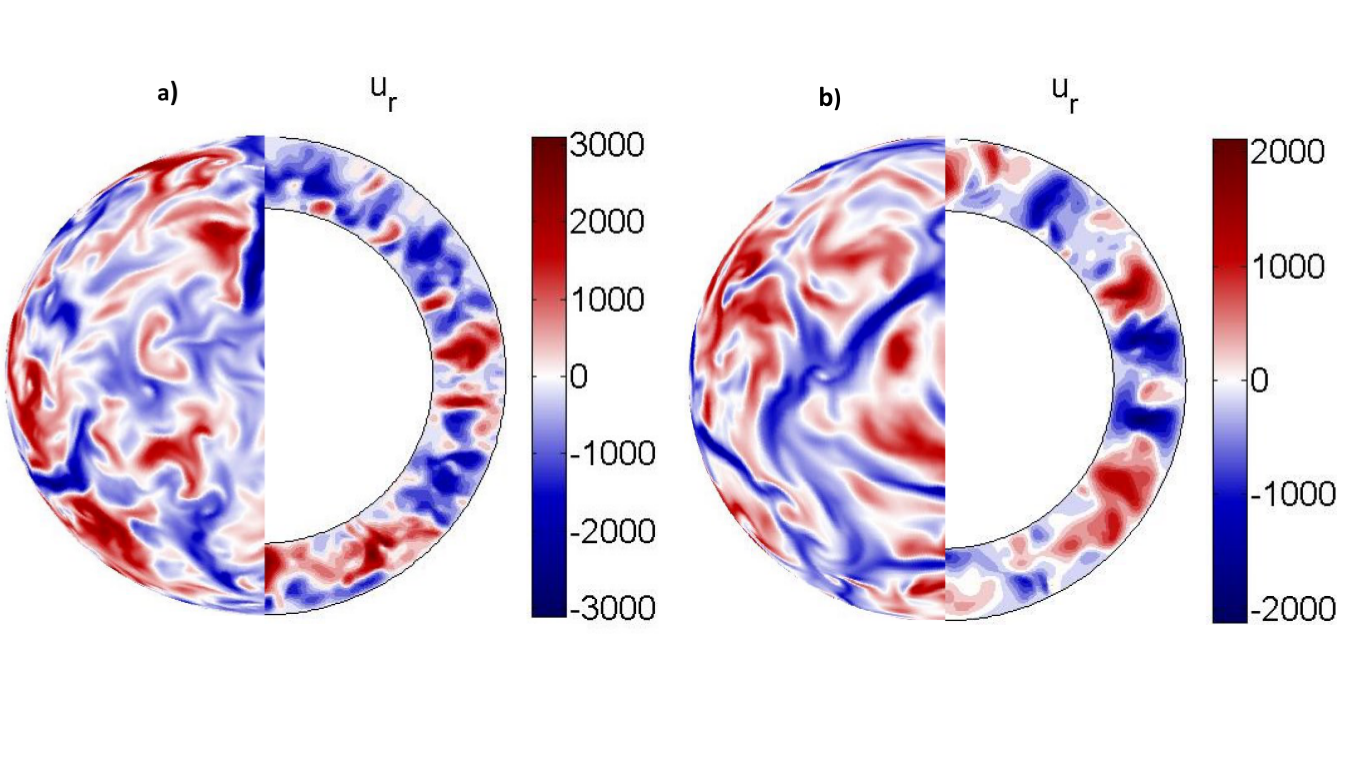} 
\caption{\label{fig_2} (Color online) Snapshot of the radial velocity $u_r$ for a) $\chi=0.1$ and b) $\chi=0.5$ and Ra$_o=10^7$. In each panel, the left part shows the $u_r$ field close to the outer radius just below the boundary layer. The right part shows the same field $u_r$ on a selected meridional plane.}
\end{figure*}

 We now  examine the qualitative and quantitative properties of our simulations, focusing on three typical cases with varying $\chi$ ($\chi=0.1$, $\chi=0.5$ and $\chi=1$) for fixed Ra$_o=10^7$.  A  simple way of visualizing the turbulent motions due to  convection  is to look at snapshots of the velocity components   $u_r$, $u_{\theta}$ or $u_{\phi}$ at a typical time after saturation of the linear instability.  Figure \ref{fig_2} shows  snapshots of $u_r$. In each panel, the left hemisphere shows the velocity field on a spherical shell  close to the upper boundary, illustrating the convective motions near the surface. The right hemisphere is a meridional slice showing the radial velocity as a function of depth and latitude, for a selected longitude. Figure \ref{fig_2}(a) is for  $\chi=0.1$, while Figure \ref{fig_2}(b) is for $\chi=0.5$. We  notice that the $\chi=0.1$ case  appears somewhat more turbulent than the $\chi=0.5$ case, as visualized by stronger eddies with a wider range of scales. 
 
\par
\begin{figure}[H]
\centering
\includegraphics[scale=0.6]{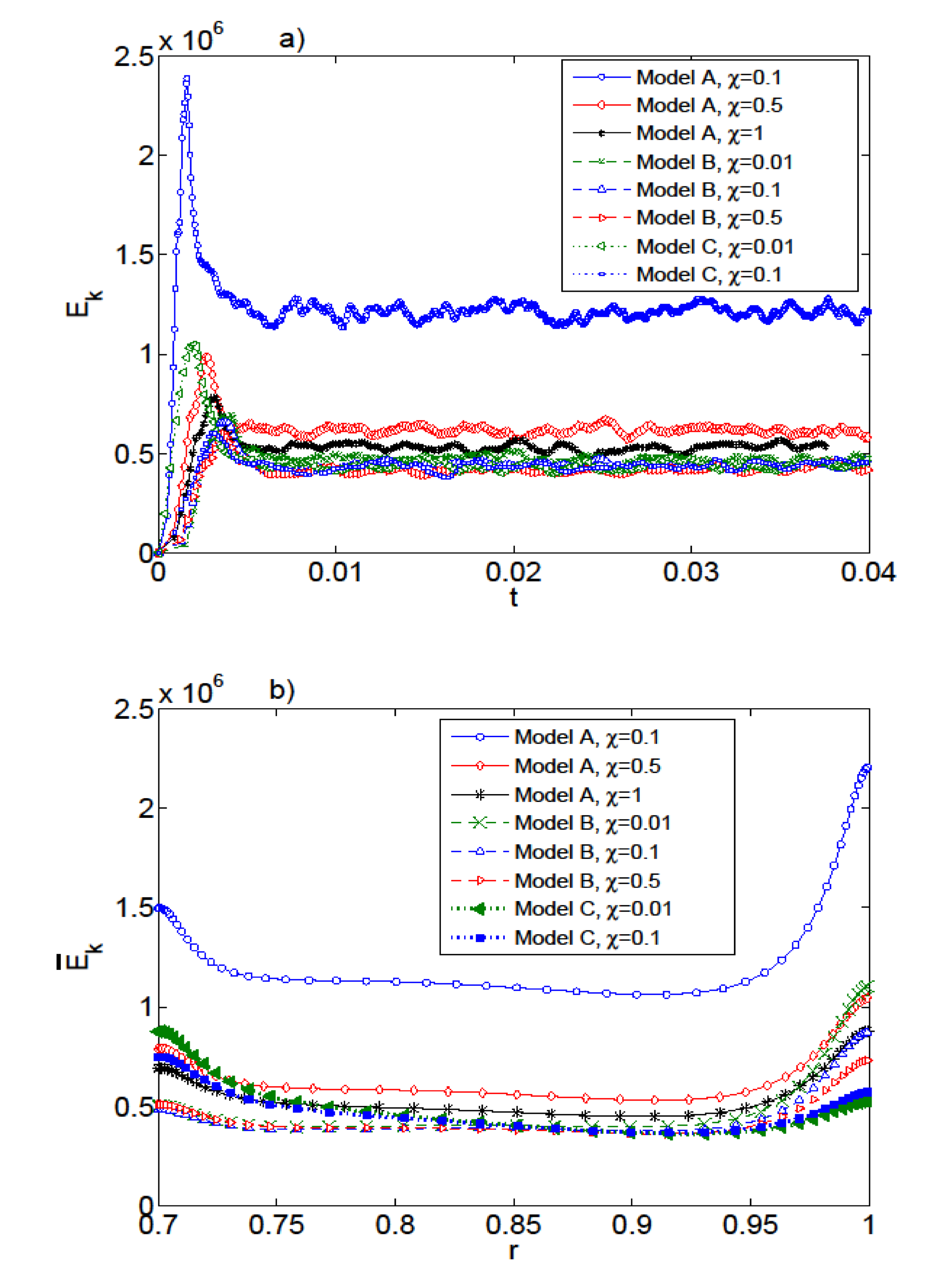}
\caption{\label{fig_3} (Color online) \textit{a)} Non-dimensional kinetic energy {\color{black}per unit volume} as a function of time for Model A (solid lines), Model B (dashed lines) and Model C (dotted lines) for Ra$_o=10^7$, and three different $\chi$.  \textit{b)} Time-averaged kinetic energy profile as a function of radius, for the same simulations.}
\end{figure}

  To see more clearly any difference among the  runs, we turn to more quantitative measures.  Figure \ref{fig_3}(a) shows the  {\color{black}kinetic energy per unit volume $E_k$} within the shell as a function of time for the Model A simulations (solid lines). We clearly observe  the  initial development of the convective instability, visible as a large spike in the interval  $t\in[0,0.01]$, followed by  its  nonlinear saturation. 
Note that   {\color{black}$E_k(t)$} reaches a stationary state  very fast  but reaching  thermal equilibrium  is a much slower process. We estimate that a simulation has reached thermal equilibrium when $\partial \Theta/\partial r |_{r= r_o} $ is statistically stationary and equal to zero. This happens around $t\approx 0.02$ for  the $\chi = 1$, $\chi = 0.5$ and $\chi = 0.1$ simulations. In all that follows, we only present the results of simulations once they have achieved thermal equilibrium.
\par Figure \ref{fig_3}(a) shows that  {\color{black}$E_k(t)$} is much larger for the  $\chi=0.1$ run than for cases with larger $\chi$, confirming our rapid visual inspection of Figure \ref{fig_2}. 
To understand why this may be the case, recall that for smaller values of $\chi$ the local Rayleigh number Ra$(r)$ increases  more with depth than for larger $\chi$  (Fig. 1). A higher Rayleigh number near the lower boundary drives convection more vigorously, which increases the overall kinetic energy.
\par
\begin{figure*}
\includegraphics[scale=0.4]{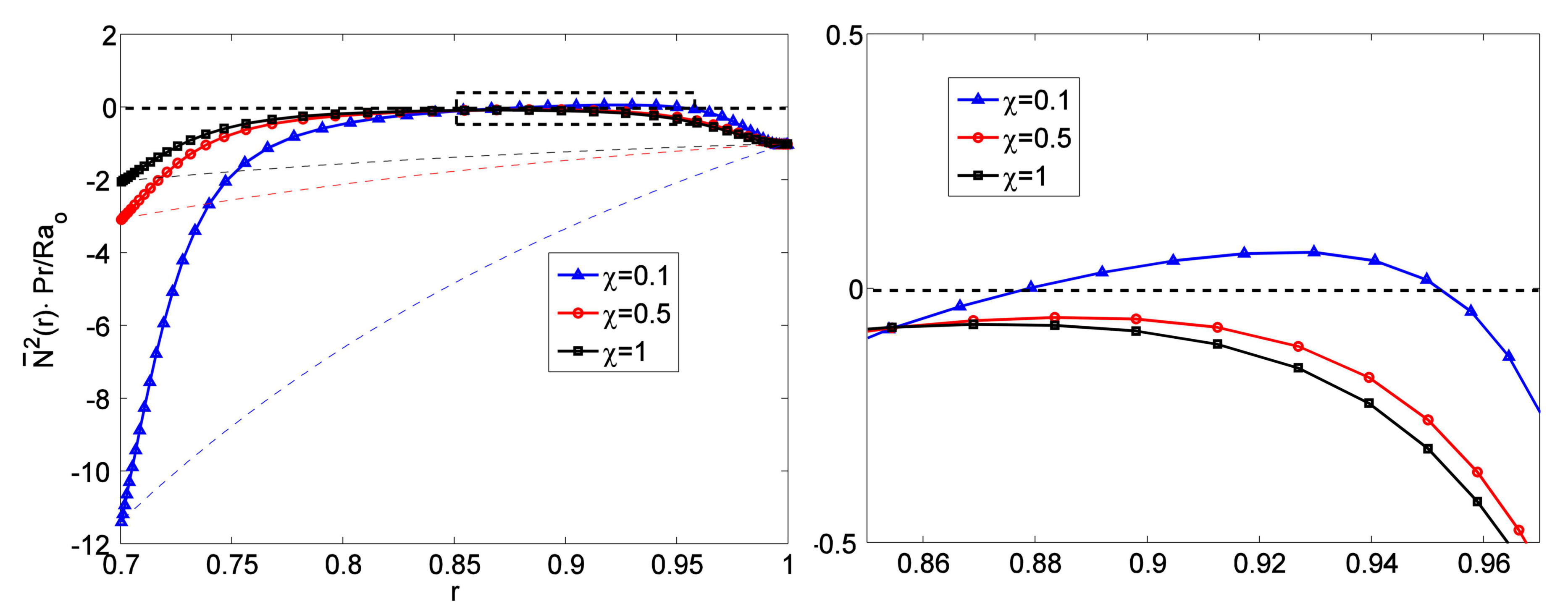}
\caption{\label{fig_4} (Color online) Profile of $\bar{N}^2$Pr$/$Ra$_o$ (solid line) compared with  $N_{\rm{rad}}^2$Pr$/$Ra$_o$ (dashed line) for $\chi=0.1$, $\chi=0.5$, $\chi=1$ and for  Ra$_o=10^7$. The right figure is a zoom-in of the dashed box in the left figure i.e.  the range where the subadiabatic region emerges.}

\end{figure*}

 Throughout the paper, we define the time- and spherical- average of a quantity as   
\begin{equation}
\bar{q}(r)=\displaystyle\frac{1}{{\color{black}4\pi(t_2-t_1)}}\int_{t_1}^{t_2}\int_0^{2\pi}\int_0^{\pi} q(r,\theta,\phi,t)\sin\theta d\theta d\phi.
\end{equation}
Figure \ref{fig_3}(b) shows  the {\color{black}non-dimensional} kinetic energy profiles $\bar{E}_k(r)$    given by
   \begin{eqnarray}
   \label{sphKE}
   {\color{black}{\bar{E}}_k(r)=\displaystyle\frac{1}{2}(\overline{{u_r^2+u_{\theta}^2+u_{\phi}^2)}}}.
   \end{eqnarray}
    The forms of these  profiles look  similar for $\chi=0.1$, $\chi=0.5$ and $\chi=1$, taking their highest value at the top of the convection zone and then decreasing inward to a plateau  from  approximately $r=0.95$  down to $r=0.75$. Below $r=0.75$, there is a small increase in the kinetic energy associated with  the inner boundary layer. As we already saw in Figure \ref{fig_3}(a), the $\chi=0.1$ case has significantly  higher kinetic energy than the other runs. Somewhat surprisingly, however, the kinetic energy is larger everywhere even though Ra$(r)$/Ra$_o$ is only larger near the inner boundary. This could be  explained by the fact that the convection in this model is a highly non-local process i.e. that stronger driving deeper down implies strong upflows and downflows throughout the domain.
\noindent  In Figure \ref{fig_4}, we plot the square of the non-dimensional  buoyancy frequency 
\begin{equation}
{\color{black}\bar{N}^2(r)} = \alpha g \left(\frac{d\bar{T}}{dr} - \frac{dT_{\rm{ad}}}{dr}\right) \frac{r_o^4}{\nu^2}=\left(\beta(r)+\frac{d\bar{\Theta}}{d r}\right)\frac{\text{Ra}_o}{\text{Pr}},
\end{equation} (solid line)
for $\chi=0.1$, $\chi=0.5$, $\chi=1$ and Ra$_o=10^7$. We also show the square of the background buoyancy frequency  ${\color{black}N^2_{\rm{rad}}(r)}=\beta(r)$Ra$_o/$Pr as a dashed line for reference. As expected, we find that  convective motions  outside the boundary layers generally mix potential temperature and drive the mean radial temperature gradient towards an adiabatic state where $\bar{N}^2\approx 0$.
However, subtle differences arise when $\chi$ decreases, which manifest themselves in two different ways.  Firstly, note that for lower $\chi$, $|N^2_{\rm{rad}}|$ is much larger, consistent with stronger convective driving.
Nonetheless, even though the  {\color{black}kinetic energy per unit volume} is larger, we see that the interior is not mixed as well for $\chi=0.1$ as for $\chi=0.5$ and $\chi=1$. Secondly, for  $\chi=0.1$, we observe the surprising emergence of a slightly subadiabatic region ($\bar{N}^2 >0$)  just below the upper boundary layer. 
\par  This remarkable behavior, {\color{black} namely} the emergence of a layer in the flow that is subadiabatic and therefore ostensibly convectively stable, only occurs for the lowest value of $\chi$ we were able to simulate.  Proceeding to lower $\chi$ to test the robustness of this observation would be an obvious path, but one that is numerically difficult.  For example, using $\chi=0.01$ would require the Rayleigh  function Ra$(r)$ to reach values of approximately $100$Ra$_o$ at the inner boundary.  Such a range is hard to resolve.     
For this reason, {\color{black}and furthermore to elicit which of the physics elements} are responsible for the unexpected emergence of a subadiabatic layer, we now switch to a different model setup (Model B).

\section{Spherical shell with a constant Rayleigh function}
\par In the   Model A simulations discussed in the previous section, both $\beta(r)$ and the local Rayleigh number Ra$(r)$ vary with depth  proportionally to one another. As a result, it is difficult to determine what may be the direct cause of some of the interesting features we observe. Thus,  we construct a second model (called  ``Model B") where $\beta(r)$ is the same as in Section II, but where Ra$(r)$ is constant across the convection zone. 
We can achieve this by  varying the thermal expansion coefficient $\alpha$ with radius  in order to compensate for the radial variation of $dT_{\rm{rad}}/dr-dT_{\rm{ad}}/dr$. Continuing to assume that  $\kappa$, $\nu$ and $g$ are constant, we now choose  $\alpha(r)$ such that    Ra$(r)$=Ra$_o$. That is,

\begin{equation}
 \displaystyle \text{Ra}(r)=-\frac{\alpha(r) g\left(\displaystyle \frac{dT_{\rm{rad}}}{dr}-\frac{dT_{\rm{ad}}}{dr}\right)r_o^4}{\kappa\nu}
=-\text{Ra}_o\displaystyle\frac{\alpha(r)}{\alpha_o} \beta(r)\equiv \text{Ra}_o
\end{equation}
as long as  ${\alpha(r)}/{\alpha_o}\equiv -{1}/{\beta(r)}{\color{black},}
$ {\color{black}where $\alpha_o  = \alpha(r_o)$}.
In this new setup, the non-dimensional  momentum equation is
\begin{equation}
\label{mom}
\displaystyle\frac{\partial{\vel}}{\partial{t}}+{\vel}\cdot{\nabla}{\vel}=-{\nabla}{ p}+\frac{\alpha(r)}{\alpha_o}\frac{\text{Ra}_o}{\text{Pr}}{\Theta}\boldsymbol{e_r}+{\nabla}^2{\vel}{\color{black},}
\end{equation}
while the thermal energy equation remains unchanged, and is given by Eq. (\ref{heq}). 
\par
\begin{figure}[H]
\centering
\includegraphics[scale=0.3]{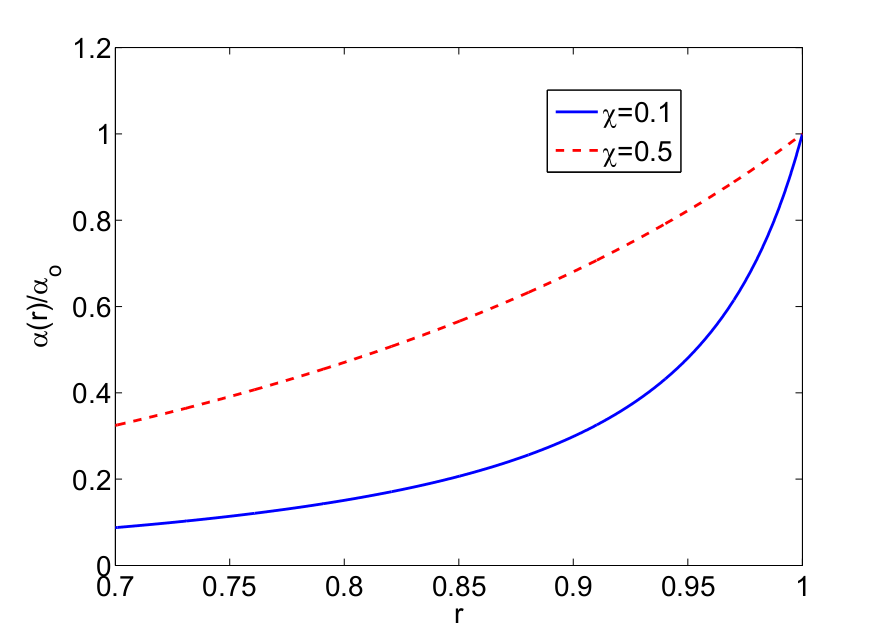}
\caption{\label{fig_5} (Color online) The dependence of $\alpha(r)$ on $\chi$ in Model B.}
\end{figure}

 
 Figure \ref{fig_5} shows the variation of $\alpha$ needed to keep  the Rayleigh function constant for our fiducial values of $\chi$.  In all cases, $\alpha(r)$ decreases with depth, and $\alpha(r_i)/\alpha_o$ is smaller for smaller $\chi$.  Physically speaking, this implies  that the effective buoyancy of fluid elements with fixed temperature perturbation $\Theta$ decreases with depth. {\color{black} Note that, in contrast with Model A, Model B now explicitly violates the conditions of use of the SVB approximation when $\chi$ is small. This will be discussed in Section VI, but lends credence to our use of the paper title ``weakly non-Boussinesq convection".}

\begin{table}[b]
\begin{ruledtabular}
\begin{tabular}{cccccccccc}
 &Model&$\chi$ &Ra$_o$
 &$N_r$ &$N_{\theta}$ &$N_{\phi}$  & $L_{max}$ & $M_{max}$ \\
\hline
\\
& A &  0.1 & $10^7$ & 250 & 402 & 480 & 268 & 134
 \\
& A &  0.5 & $10^7$ & 220 & 346 & 384 & 230 & 120 
 \\
& A &  1 & $10^7$ & 220 & 346 & 384 & 230 & 120 
 \\
& B & 0.001 & $10^7$ & 200 & 288 & 320 & 192 & 96 
\\
& B &  0.01 & $10^6$ & 200 & 192 & 192 & 128 & 64
\\ 
& B &  0.01 & $10^7$ & 200 & 288 & 320 & 192 & 96 
\\ 
& B &  0.01 & $10^8$ & 300 & 516 & 640 & 344 & 172
\\ 
& B &  0.1 & $10^7$ & 200 & 288 & 320  & 192 & 96 
\\ 
& B &  0.1 & $10^8$ & 300 & 516 & 640  & 344 & 172
\\ 
& B &  0.5 & $10^7$ & 200 & 288 & 320  & 192 & 96 
\\ 
& C &  0.01 & $10^7$ & 200 & 288 & 320 & 192 & 96 
\\
& C &  0.1 & $10^7$ & 200 & 288 & 320 & 192 & 96 
\end{tabular}
\end{ruledtabular}
\caption{\label{table1} Table with all the different model configurations and the  input parameters used in each case. The resolution is provided  both in number of equivalent meshpoints $N_r$, $N_{\theta}$, $N_{\phi}$, as well as in the number of spherical harmonics used in the horizontal directions,  $L_{max}$ and $M_{max}$.} 
\end{table}

\par We now compare the convective dynamics of the Model A and B  setups in order to try and understand the respective roles of Ra$(r)$ and $\beta(r)$ in driving convection and mixing. To do so, we have run  numerical simulations using Model B for three different values of $\chi$, for a fixed Ra$(r)=$Ra$_o$=$10^7$. In these constant Rayleigh function runs, we were able to achieve values of $\chi$ down to  $0.001$. 
\par  Figure \ref{fig_3} compares the energetics of Model A (solid lines) and Model B (dashed lines) runs. In Figure \ref{fig_3}(a), we observe that   the saturation level of the  {\color{black}kinetic energy per unit volume $E_k$ varies much less with $\chi$ in Model B than in Model A.} This might be expected since both Ra and Pr are now constant at all radii and in all configurations of Model B presented. 
In Figure \ref{fig_3}(b), we see that the  {\color{black}kinetic energy} profiles $\bar{E}_k(r)$ of the various Model B runs almost coincide in the bulk of the convection zone, showing that $\beta$ alone does not influence this quantity much.   
\par
\begin{figure}[H]
\centering
\includegraphics[scale=0.3]{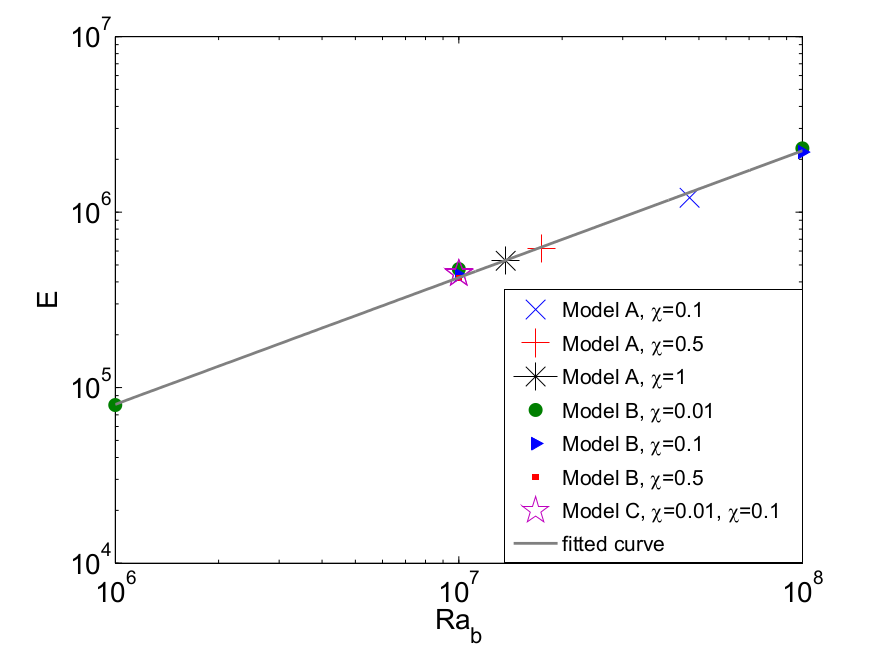}
\caption{\label{fig_6} (Color online) Mean kinetic energy $E$ versus bulk Rayleigh number ${\color{black}{\rm{{Ra}}}_b}$ for all the Models. Configurations with the same bulk Rayleigh number have approximately the same kinetic energy. The straight line is a fit to the data, with  $E = (3.7\pm 2.6){\color{black}{\rm{{Ra}}}_b}^{0.72\pm 0.04}$. }
\end{figure}

  Figure \ref{fig_6}  shows  the mean kinetic energy ${E}$ (i.e. the time average of  $E_k$)  against the bulk Rayleigh number  {\color{black} defined as}
\begin{equation}
\label{Ranum}
\displaystyle{\color{black}{\rm{{Ra}}}_b}=\frac{\int_{r_i}^{r_o}{\rm{Ra}}(r)r^2 dr}{\int_{r_i}^{r_o}r^2dr}.
\end{equation}  
We see that the mean kinetic energy  depends solely on the bulk Rayleigh number {\color{black}(for fixed Pr and $r_o-r_i$)} and not on  the  setup used or on the value of $\chi$. This is a very interesting finding, since it illustrates that the mean kinetic energy is model-independent and can be predicted as long as  the bulk Rayleigh number ${\color{black}{\rm{{Ra}}}_b}$ of the problem is known.  Fitting the available data, we find that $E = (3.7\pm 2.6){\color{black}{\rm{{Ra}}}_b}^{0.72\pm 0.04}$. {\color{black} Note that we expect the exponent to be universal, but, the prefactor likely depends on Pr or on $(r_o-r_i)$.}

\par
\begin{figure}[H]
\centering
\includegraphics[scale=0.3]{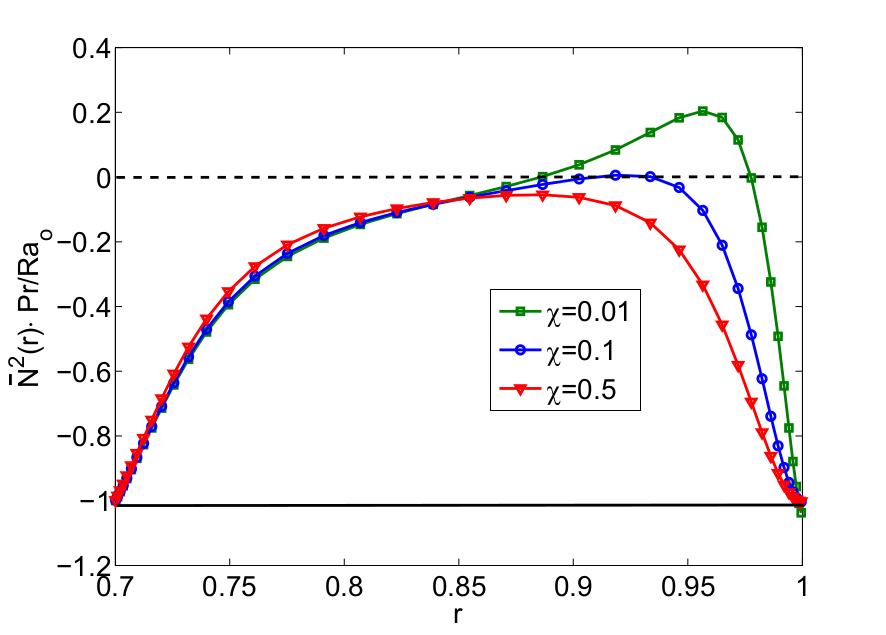}
\caption{\label{fig_7} (Color online) $\bar{N}^2(r)$Pr$/$Ra$_o$ profile compared with $N^2_{\rm{rad}}(r)$Pr$/$Ra$_o\equiv -1$ (solid black line) for different values of $\chi$ and Ra$_o=10^7$ ({\color{black}all runs are using} Model B). Note how the subadiabatic region becomes much more pronounced for lower $\chi$.}
\end{figure}

 In order to compare the efficiency of mixing in this new system, we again look at the square of the  non-dimensional buoyancy frequency, defined for Model B as 
\begin{equation}
\bar{N}^2(r)=\frac{\alpha(r)}{\alpha_o}\left(\beta(r)+\frac{d\bar{\Theta}}{dr}\right)\frac{\text{Ra}_o}{\text{Pr}}.
\end{equation}
 In Figure \ref{fig_7}, we plot  $\bar{N}^2(r)$Pr/Ra$_o$ compared with the background  $N^2_{\rm{rad}}$Pr/Ra$_o=(\alpha(r)/\alpha_o)\beta(r)$ for  Model B. Note that by construction in this setup  $N^2_{\rm{rad}}$Pr$/$Ra$_o=-1$ regardless of $\chi$. We see that as $\chi$ decreases, $\bar{N}^2$  increases and for $\chi\leq 0.1$ a subadiabatic region does indeed emerge as  in Model A.
This unusual effect  is much more pronounced at   $\chi=0.01$. Overall, this conclusively shows that the appearance of the subadiabatic region is not model-dependent, but instead, a fairly generic property  of  simulations that combine mixed temperature boundary conditions with varying superadiabaticity. 

\par
\begin{figure}
\centering
\includegraphics[scale=0.3]{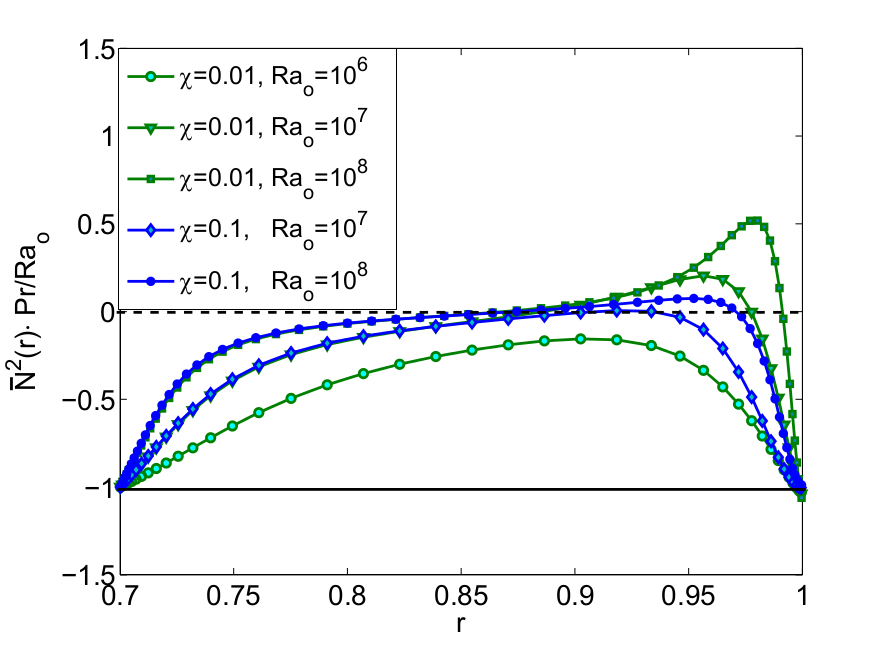}
\caption{\label{fig_8} (Color online) $\bar{N}^2(r)$Pr$/$Ra$_o$ profile for Model B for different values of $\chi$ and Ra$_o$. The solid black line indicates the background $N^2_{\rm{rad}}(r)$Pr$/$Ra$_o=-1$.}
\end{figure}

 In order to determine more precisely how the emergence of a subadiabatic region depends on the model parameters, we ran additional  simulations at Ra$_o=10^6$ and  Ra$_o=10^8$ for  $\chi=0.01$, as well as a simulation with Ra$_o=10^8$ for  $\chi=0.1$.  Figure \ref{fig_8} shows the square of the buoyancy frequency profiles for these comparative runs. We observe that, for a given value of $\chi$, there is a threshold value of Ra$_o$ above which the subadiabatic region appears, and that the size and subadiabaticity of that region increases with Ra$_o$ beyond that threshold. For fixed Ra$_o$ we see a similar behavior with decreasing  $\chi$. These considerations suggest that the subadiabatic layer appears only for sufficiently vigorous convection (high Rayleigh number) and/or in systems with sufficiently large radial variations in the background superadiabaticity {\color{black}(here generated by low $\chi$)}. 

\par
\begin{figure*}
\includegraphics[scale=0.5]{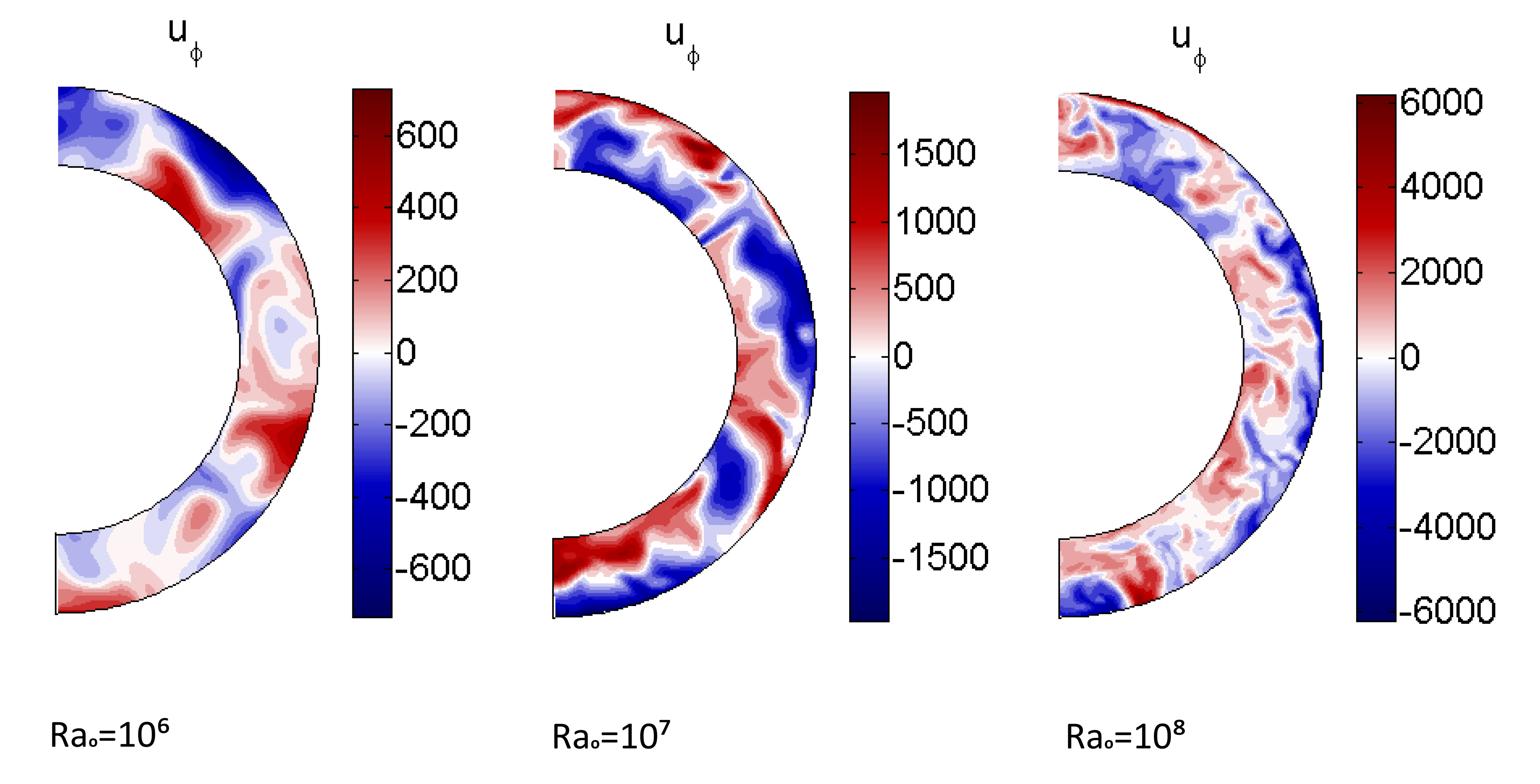} 
\caption{\label{fig_9} (Color online)
 Snapshots of $u_{\phi}$ in a selected meridional slice for Model B when $\chi=0.01$ and for three different Ra$_o$. As we increase the Rayleigh number, the convective eddies are more pronounced and the turbulent motions are more intense.}
\label{usRa}
\end{figure*}

\begin{figure}[H]
\centering
\includegraphics[scale=0.3]{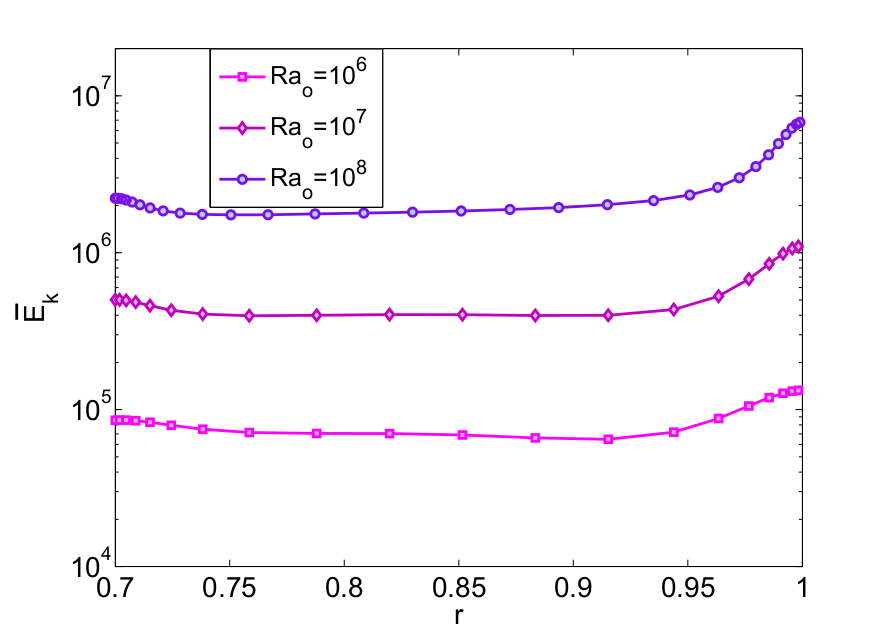}
\caption{\label{fig_10}
 (Color online) Time-averaged kinetic energy profile $\bar{{E}}_k(r)$  for Model B, $\chi=0.01$, and for three different values of Ra$_o$.}
\end{figure}

 Interestingly, convection appears unaffected by the emergence of the subadiabatic layer, and proceeds as if it did not exist. This can be seen both in snapshots of the velocity field (Figure \ref{fig_9}) and in the kinetic energy profiles as a function of radius (Figure \ref{fig_10}). Figure \ref{fig_9} shows snapshots of $u_{\phi}$ as a function of radius and latitude, for a fixed longitude, for $\chi = 0.01$ and the three different Rayleigh numbers used in that case.  As the Rayleigh number increases, the convective eddies are more pronounced and the  turbulent motions are apparently  stronger. However, none of the simulations show any obvious indication of a non-convective or ``dead" zone due to the subadiabatic layer (which is present for the Ra$_o = 10^7$ and Ra$_o=10^8$ cases).  The same can be seen more quantitatively in Figure \ref{fig_10}, which shows the  kinetic energy profiles for the same three cases (Ra$_o = 10^6$, $10^7$ and $10^8$, and $\chi =0.01$).  As in Model A, we find that they have roughly the same shape, but that the  {\color{black}kinetic energy} increases with Ra$_o$. Crucially, there is no sign of any dip in the kinetic energy profiles at the locations of the subadiabatic layers in the Ra$_o = 10^7$ and Ra$_o = 10^8$ runs, which proves that convection is efficient everywhere across the shell. All the above provide  strong indications that convection in these models is a very non-local process.

\section{Interpretation of the results}
Having established that the emergence of a subadiabatic layer is a robust phenomenon in these models, we now proceed to explain the observed dynamics more quantitatively. As we shall demonstrate, the phenomenon is directly related to the use of mixed temperature boundary conditions, and {\color{black} is facilitated by} the presence of a strongly-varying background superadiabaticity which act together to create strong asymmetries between  upflows and downflows. 
\par We note that in traditional Boussinesq Rayleigh-B\'{e}nard convection between parallel plates with fixed temperature boundary conditions (BRBC thereafter),  an initially superadiabatic mean temperature profile becomes relaxed via convective motions to a state where upper
and lower superadiabatic boundary layers are joined by an adiabatic interior. Upflows and downflows are driven by buoyancy forces in the boundary layers, and interact non-linearly in the bulk of the fluid, mixing it towards an adiabat. The intrinsic up-down symmetry of BRBC implies that upflows and downflows contribute equally to the upward heat transport. As we now demonstrate, our findings here are  very  different. 
\par
\begin{figure}[H]
\centering
\includegraphics[scale=0.6]{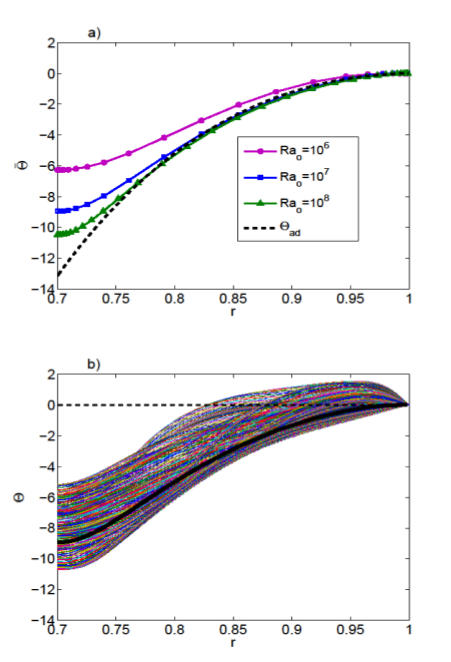}
\caption{\label{fig_11} (Color online)  a) {\color{black}Spherically-} averaged temperature perturbation profile $\bar{\Theta}(r)$, compared with $\Theta_{\rm{ad}}(r)$ (see main text for detail) (dashed black line) for $\chi=0.01$, and three different values of Ra$_o$  for Model B runs. b) Temperature perturbation {\color{black} profiles (thin colored lines)} $\Theta(r,\theta,\phi,t)$ at a fixed time $t$ and fixed longitude $\phi$, for varying $\theta$ and $r$ compared with the spherical average {\color{black}$\bar{\Theta}(r)$ profile (thick black line)} for the $\chi=0.01$, and Ra$_o=10^7$ Model B run.}
\end{figure}

  In Figure \ref{fig_11}(a), we show  the  temperature perturbation profiles
$\bar{\Theta}(r)$ in Model B runs with $\chi=0.01$, and Ra$_o=10^6$, Ra$_o=10^7$ and Ra$_o=10^8$. This  quantity is proportional to the term (Ra$/$Pr)$\bar{\Theta}(r)$ which is the time- and  spherically-averaged non-dimensional buoyancy force in the statistically stationary state (see  Equation (\ref{mom})). We notice that $\bar{\Theta}$ is  negative almost everywhere so the average buoyancy force is   downwards. That is quite different from what happens in the BRBC case (with or without adiabatic temperature gradient), where $\bar{\Theta}$ would be found to be positive in the bottom half of the domain, and negative in the top half.  This raises two questions:  (1) why is $\bar{\Theta}$  almost entirely negative in our setup, and (2) how are the  upflows driven if the average buoyancy force is downwards?\\
To answer the first question, we look at the behavior of an individual fluid parcel. Ignoring thermal diffusion,  the evolution of the temperature within the parcel is given by the Lagrangian derivative $D\Theta / Dt = - \beta u_r$, which can be re-written as  ${\color{black} D\Theta/Dr} = - \beta$ since the radial velocity of the parcel is $u_r = D r / Dt$. The temperature of the parcel at the boundary {\color{black}is given by  $\Theta(r_o) = 0$}, therefore by integration inwards, the temperature within downward flowing parcels is $\Theta_{\rm{ad}}(r)=-\int_{r_o}^r \beta(r') dr'$. This  quantity  represents the temperature perturbation within a fluid parcel at radius $r$ that is moving downward adiabatically and without any mixing with its surroundings.  Comparing $\Theta_{\rm{ad}}(r)$ to the mean temperature perturbation profiles  $\bar{\Theta}(r)$ of Model B simulations with varying Ra$_o$ in Figure \ref{fig_11}(a), we see that they  coincide  throughout most of the domain except when approaching the inner boundary. This suggests that indeed the fluid parcels travel downwards more or less adiabatically, and that is what controls $\bar{\Theta}$ over much of the spherical shell. 

\par In answer to the second question, there are two possibilities: either the mean hides information about the upflows and perhaps the return flows from the lower boundary are very rare but strong, arising from  the tail of the distribution of the temperature perturbations $\Theta$ around the mean $\bar{\Theta}$, {\color{black}or  pressure} dominates over buoyancy and pushes the parcels back upwards. 
 Figure \ref{fig_11}(b) shows   $\Theta$  profiles at specific {\color{black}example} locations (fixed longitude {\color{black}but various latitudes}) and a specific time, and indicates that there really are no fluid parcels with positive $\Theta$ anywhere in the lower part of the domain. We thus conclude that the upflows are not buoyantly driven, and must be pressure dominated.

\par
\begin{figure}[H]
\centering
\includegraphics[scale=0.3]{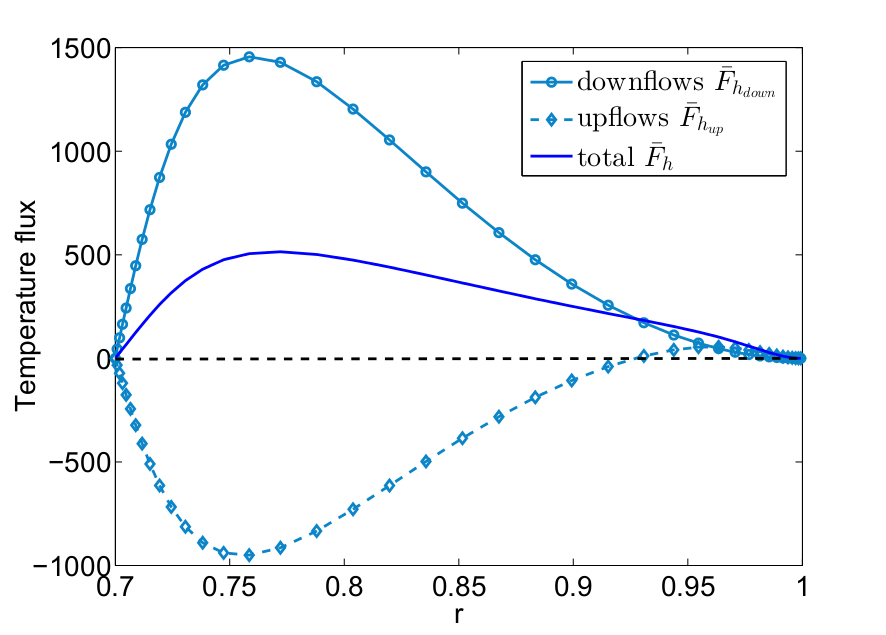}
\caption{\label{fig_12} (Color online) Downward and upward temperature fluxes (along with their sum which gives the total  turbulent temperature flux) for a Model B run with $\chi=0.01$, and Ra$_o=10^7$.}
\end{figure}

\begin{figure}[H]
\centering
\includegraphics[scale=0.6]{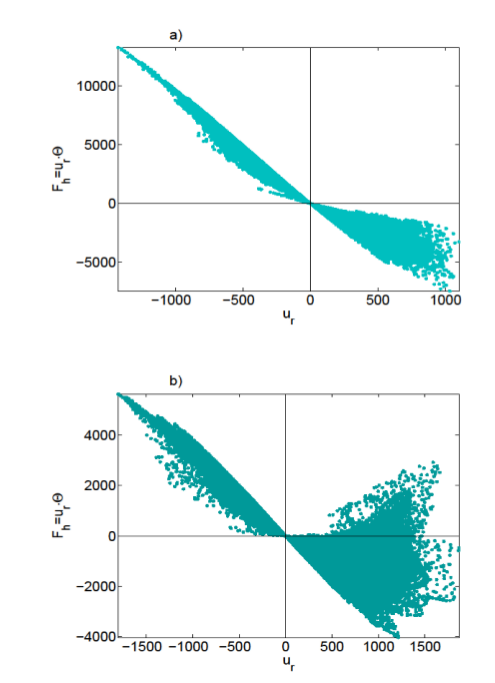}
\caption{\label{fig_13} (Color online) A typical scatter plot of the  turbulent temperature flux against the radial velocity for a Model B run with $\chi=0.01$ and Ra$_o=10^7$ calculated  \textit{a)} close to the inner radius at $r\approx 0.75$ and  \textit{b)} in the middle of the domain at $r\approx 0.85$.}
\end{figure}

 This idea can be confirmed by looking at the turbulent temperature fluxes and the respective contributions from the upflows and downflows directly. 
  In quintessential convection such as  BRBC, the direction of the force acting on a fluid parcel is given by the sign of $\Theta$. As a result, aside from short transients, there is a very strong correlation between the sign of the temperature perturbation and the sign of the vertical velocity of the fluid parcel (with  $\Theta>0$ corresponding to $u_r>0$, and $\Theta<0$ corresponding to $u_r<0$). Therefore,  in both  cases,  one might expect the turbulent (or convective) temperature flux $F_h=u_r\Theta$  to be positive, if motion is due to buoyancy. With this in mind, we therefore examine the contributions from upflows and  downflows to the total turbulent temperature flux
separately. We investigate by looking firstly at the mean flux (Fig. \ref{fig_12}) and then secondly at  the pointwise flux (Fig. \ref{fig_13}),  for a characteristic Model B simulation with $\chi=0.01$, and Ra$_o=10^7$ at a typical time when the system is in a statistically stationary and {\color{black}thermally relaxed} state. 
 In Figure \ref{fig_12}, we plot $\bar{F}_h$ together with the corresponding mean turbulent temperature flux carried by the upflows only, $\bar{F}_{up}$ (given by $\bar{F}_h$ using only those points where  $u_r>0$) and by the downflows only, $\bar{F}_{down}$ (given by $\bar{F}_h$ using only those points where  $u_r<0$). 
 We notice that the average flux carried by the   downflows is positive, which means that they transport  relatively cold material downward as expected. However, the average flux carried by the upflows is negative, indicating that they are carrying cold material up, contrary to expectations.  In standard BRBC for instance, both downflows and upflows would on average transport heat upwards, i.e.  $\bar{F}_{h}$, $\bar{F}_{up}$ and $\bar{F}_{down}$ would all be positive, with downflows carrying relatively cold fluid and upflows carrying relatively warm fluid.   

\par To check whether this odd behavior of the upward fluxes is true only {\color{black} in the} average, or applies to all fluid parcels, we now look at the pointwise turbulent temperature flux, $F_h = u_r\Theta$. Figure \ref{fig_13} shows the pointwise flux at every point on two spherical shells, located close to the inner boundary (at $r \approx 0.75$, in Figure \ref{fig_13}(a)) and in the bulk of the domain (at $r \approx 0.85$, in Figure \ref{fig_13}(b)) respectively at a representative time. 
 As mentioned above, one might more normally (in BRBC for example) expect that nearly all  points would lie in the two upper quadrants.    
Figures \ref{fig_13}(a) and \ref{fig_13}(b) confirm our findings from Figure \ref{fig_12}, but provide more detail. While the downflows all appear to be working productively at transporting
heat upwards (i.e. cold downwards; upper left quadrant), it is clear that a considerable number of upflows are working counter-productively at both depths. At $r \approx 0.85$, in the middle of the domain,  some of the upflows work productively to transport heat upwards (upper right quadrant) but many more work counter-productively
to transport heat downwards (lower right quadrant), leading to the negative mean flux in the upflows seen in Figure \ref{fig_12}. Closer to the inner boundary all of the upflows are counterproductive to heat transport.

All of these results point to the same conclusion, namely that the upflows are not buoyantly driven {\color{black}in the lower part of} the domain. Since the only other force in the system is the pressure gradient, we conclude that the upflows must be pressure-driven. 
{\color{black} The dynamic pressure} is really only a manifestation of the divergence condition in the Boussinesq approximation, thus, another equivalent interpretation is that the upflows are merely an inertial continuation of the downflows which are forced to turn around at the lower boundary. Indeed, if a downflow driven by its negative temperature perturbation  simply ``rebounds"  off the lower boundary without changing its heat content, it becomes a  counter-productive upflow, i.e. fluid  moving upwards but with the same negative temperature perturbation. It is perhaps surprising  that this result persists so high into the shell. This appears to be a feature of low $\chi$, low Pr and high Ra$_o$  convection.

\par The above results are implicitly related to the choice of mixed boundary conditions for  $\Theta$. The  fixed flux at the inner boundary is a source of strong asymmetry in the dynamics of the problem since it allows the temperature perturbations $\Theta$ to be  negative there, roughly following the adiabatic profile $\Theta_{\rm{ad}}(r)$. In a system with fixed temperature conditions on the other hand, $\Theta$ and therefore $\bar{\Theta}$ would be forced to be zero at both bottom and top  boundaries, and the system would be much more symmetric (though not perfectly because of the sphericity and  the non-zero constant adiabatic temperature gradient which are additional sources of asymmetry). This would then guarantee the existence of temperature perturbations of both signs near the inner boundary and therefore some buoyantly-driven upflows there. 

\par
\begin{figure}[H]
\centering
\includegraphics[scale=0.3]{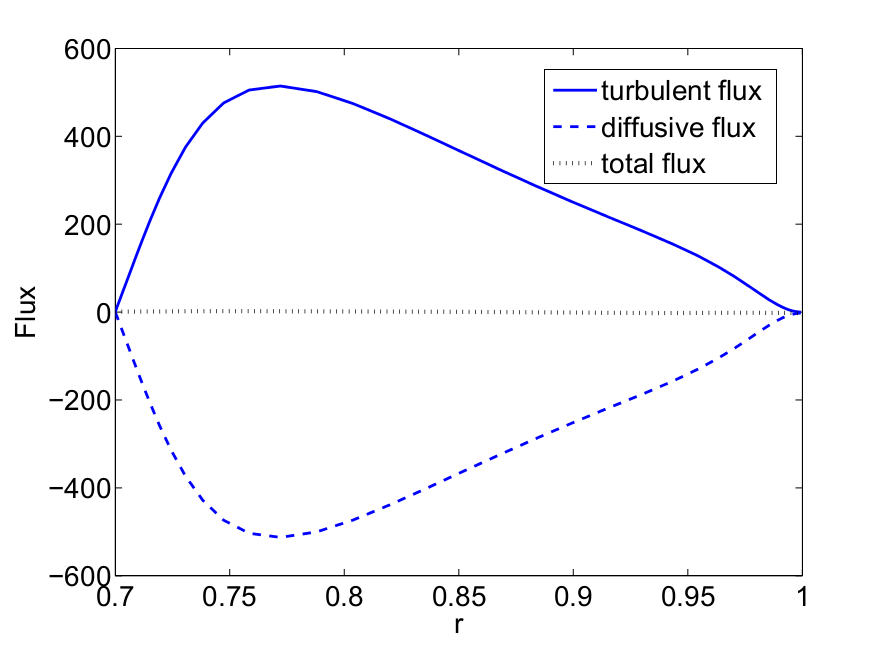}
\caption{\label{fig_14}
 (Color online) The time- and spherically- averaged turbulent and diffusive contributions to the perturbed temperature flux and their sum, for a Model B run with   $\chi=0.01$ and Ra$_o=10^7$.}
\end{figure}

 The use of mixed boundary conditions has a second very important impact on the convective dynamics, namely that the total perturbed temperature flux through the system (turbulent + diffusive) must be equal to that at the inner boundary, and thus zero everywhere (Fig. \ref{fig_14}). Non-dimensionally, this is expressed as \begin{equation}
\label{fluxeq}
\displaystyle
\bar{F}_h-\frac{1}{\rm{Pr}}\frac{d\bar{\Theta}}{d r }=0.
\end{equation}
 In thermal equilibrium, the diffusive and non-diffusive contributions to the perturbed temperature  flux must therefore cancel out exactly. The magnitude of the temperature perturbations depends on $\chi$ (through the increasingly negative values of $\Theta_{\rm{ad}}$ as $\beta(r)$ decreases rapidly with $r$
for low $\chi$) as in Figure \ref{fig_11}(a). Furthermore, the rms velocity of the convective eddies increases substantially with Ra$_o$ (see Figure \ref{fig_6}). Thus for low $\chi$ and high Ra$_o$, the turbulent flux increases and the diffusive flux of the temperature
perturbations  must follow accordingly. This is crucial, and causes the emergence of the subadiabatic layer {\color{black} in our simulations} as follows. \\
Using Equation (\ref{fluxeq}) we can re-write $\bar{N}^2$ as 
\begin{equation}
\label{N2eq}
\displaystyle \bar{N}^2(r)=\frac{\alpha(r)}{\alpha_o}(\beta(r)+{\rm{Pr}} \bar{F}_h)\frac{\rm{Ra}_o}{\rm{Pr}},
\end{equation}
 where we carefully note that $\beta(r)<0$ while $\bar{F}_h>0$. Since  $\bar{F}_h$ increases monotonically with increasing Ra$_o$  (because of the increase in  $u_{rms}$)  or decreasing $\chi$  (because $\Theta_{rms}$ is larger), there exists a region of  parameter space where $(\beta(r)+{\rm{Pr}} \bar{F}_h)$ becomes positive, at least somewhere within the shell, leading to a positive $\bar{N}^2(r)$.

\begin{figure}[H]
\centering
\includegraphics[scale=0.3]{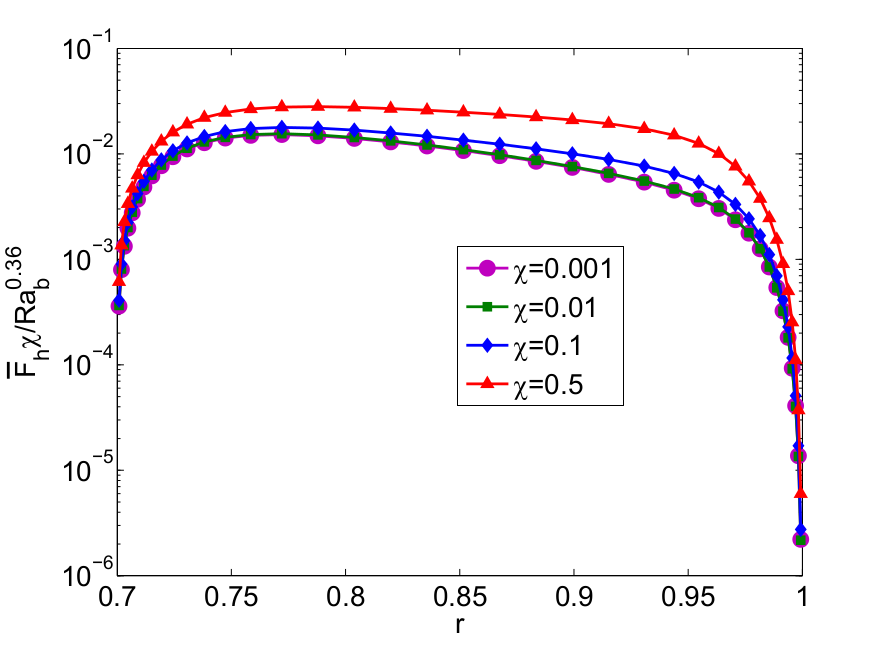}
\caption{\label{fig_15}
 (Color online) $\bar{F}_h\chi/$Ra$_b^{0.36}$ for Model B, $\chi=0.001$, $\chi=0.01$, $\chi=0.1$ and $\chi=0.5$, and  Ra$_b=10^7$.}
\end{figure}

Using the results we have obtained so far, we can in fact provide an order-of-magnitude estimate for $\bar{F}_h$ as a function of $\chi$ and of the bulk Rayleigh number ${\color{black}{\rm{{Ra}}}_b}$ given in   (\ref{Ranum}). 
The typical amplitude of the temperature perturbations $\Theta_{rms}$ can be estimated from $\Theta_{\rm{ad}}$, which is proportional to $1/\chi$ for low enough $\chi$ . The rms velocity of the flow $u_{rms}$ can be estimated from ${\color{black}{\rm{{Ra}}}_b}$ using $u_{rms} = \sqrt{2{E}}${\color{black}. In}
 Section III, we found that $E = (3.7\pm 2.6){\color{black}{\rm{{Ra}}}_b}^{0.72\pm 0.04}$, so $u_{rms}$ approximately scales as ${\color{black}{\rm{{Ra}}}_b}^{0.36}$.
Combining these two estimates suggests that the  turbulent temperature flux should scale as ${\color{black}{\rm{{Ra}}}_b}^{0.36}/\chi$ for low enough $\chi$. 
 In Figure \ref{fig_15}, we plot $\bar{F}_h\chi/$Ra$_b^{0.36}$ versus $r$ for Model B runs at Ra$_o=10^7$ and for four different values of $\chi$. The predicted scaling seems to work well for $\chi\le 0.1$. We conclude that the emergence of a subadiabatic layer  is a generic result which occurs for large bulk Rayleigh numbers {\color{black} and/or} low values of $\chi${\color{black}, as we have found in our simulations. Note that the scaling $\bar{F}_h\sim {\color{black}{\rm{{Ra}}}_b}^{0.36}/\chi$ suggests that the subadiabatic layer could in fact appear even when $\chi=1$ provided ${\color{black}{\rm{{Ra}}}_b}$ is large enough. In that sense, it might even be realized in the limit of a Cartesian RBC system as long as mixed thermal boundary conditions are used, though the Rayleigh number may need to be extremely large in that limit to exhibit the desired dynamics.}

\section{A more solar $\beta(r)$ profile: Setup and numerical results}
\begin{figure}[H]
\centering
\includegraphics[scale=0.3]{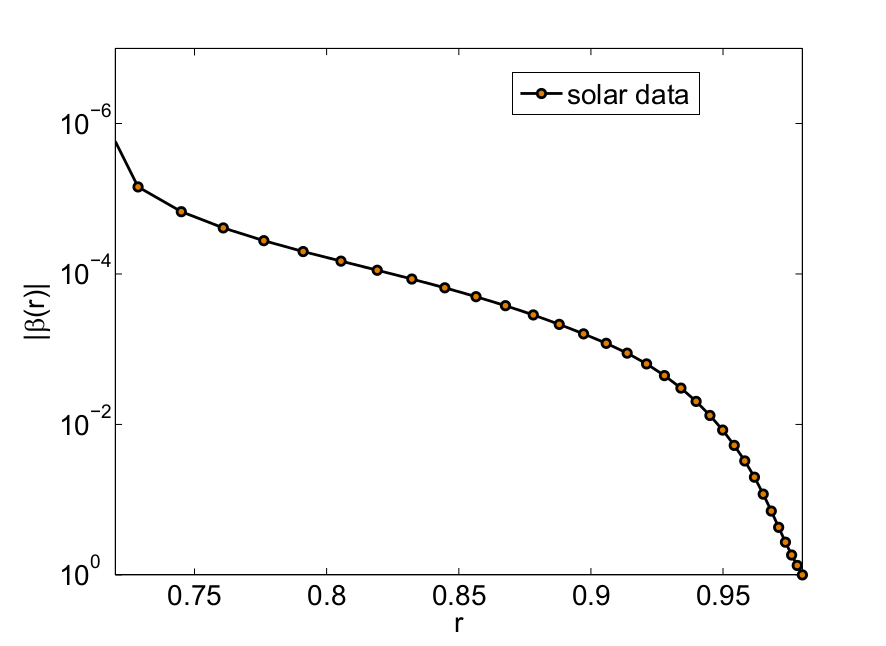}
\caption{\label{fig_16}
 (Color online) The function $|\beta(r)|$ according to Model S \cite{ModelS}.}
\end{figure}

Until now we have used a profile for $\beta(r)$ dictated by the geometry and the boundary conditions of our model setup.
To see whether our findings have any bearing on the dynamics of convection in stars, we now compute the {\color{black} equivalent} $\beta(r)$ profile from a standard solar model (Model S, \cite{ModelS}). To do so, we evaluate the difference between $dT_{\rm{rad}}/dr=-3\kappa\rho L/(64\pi r^2\sigma T^3)$ (where the Model S is used to extract   the density $\rho$, the luminosity $L$, the temperature $T$, and the opacity $\kappa$, and where $\sigma$ is the Stefan-Boltzmann  constant), and the adiabatic temperature gradient  $dT_{\rm{ad}}/dr=-g/c_p$. The results are shown in Figure \ref{fig_16}. We see that, by contrast with the models we have been using so far,  $|\beta|$ \textit{decreases} inwards instead of increasing inwards.
\par
\begin{figure}[H]
\centering
\includegraphics[scale=0.3]{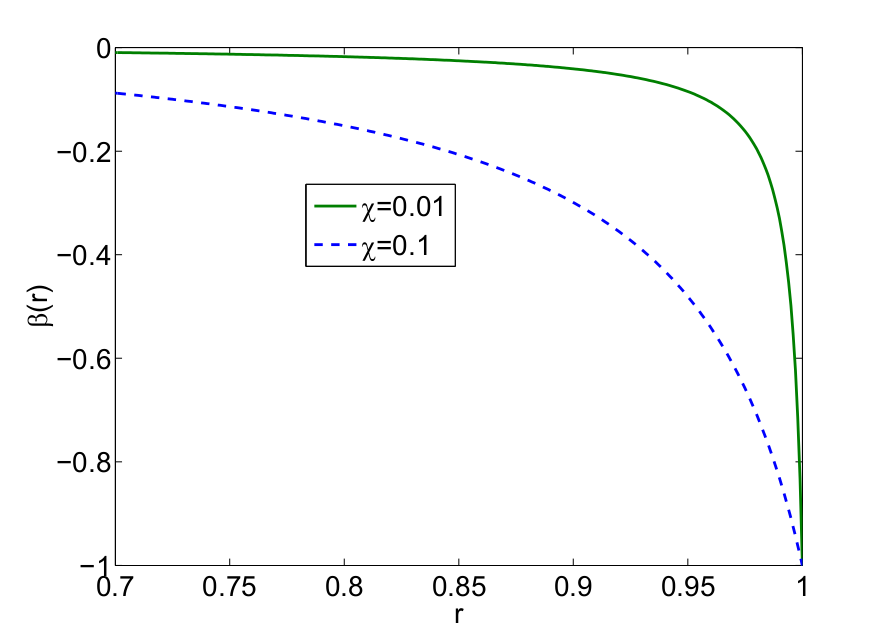}
\caption{\label{fig_17}
 (Color online) The different $\beta(r)$ profiles for $\chi=0.01$ and $\chi=0.1$ for Model C.}
\end{figure}

 In the light of this information, we conduct a final set of numerical experiments where we construct a more solar-like  profile for  $\beta(r)$ choosing $\beta(r)=\chi/(1-\chi-(1/r^2))$ in order to ensure that $|\beta(r)|$ decreases inward, and letting $\alpha(r)/\alpha_o=-1/\beta(r)$ as before to have Ra$(r)=$Ra$_o$. {\color{black} Note that in this model, $\chi$ does not have the same physical meaning as in Equation (\ref{chi}), but it is merely used as a parameter to describe a family of functions $\beta(r)$ with a  ``solar-like" profile.} Figure \ref{fig_17} illustrates the $\beta(r)$ functions  thus created  for two different  values of $\chi$. Note that $\beta$ lies in the range (0,1] but {\color{black} crucially} the ratio of the inner to outer values is large and approximately equal to $11$ for $\chi=0.1$ and $105$ for $\chi=0.01$.
\par
\begin{figure}[H]
\centering
\includegraphics[scale=0.3]{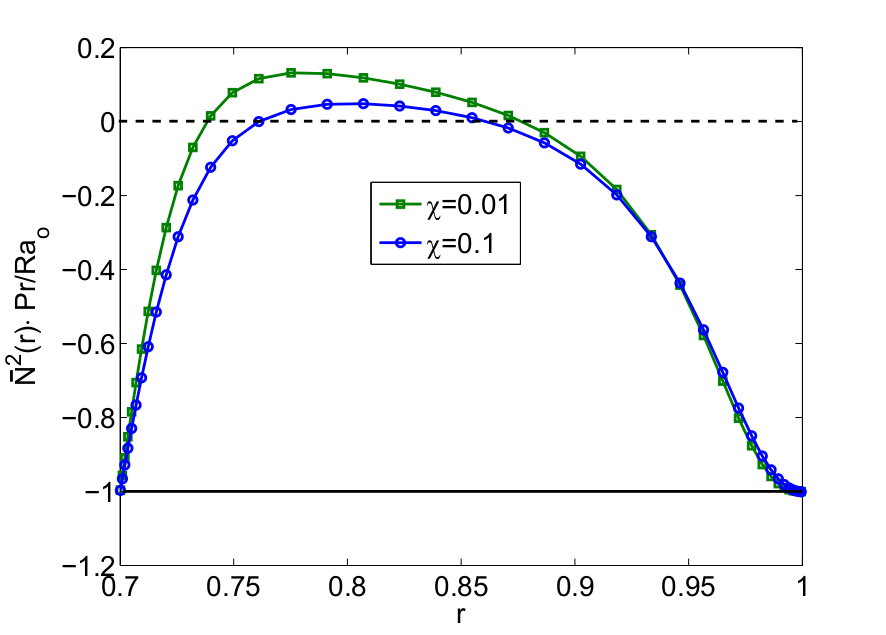}
\caption{\label{fig_18}
 (Color online) The square of the buoyancy frequency for Model C, for  two different values of $\chi$, and for Ra$_o=10^7$.}
\end{figure}

 We have run two simulations, for two different values of $\chi$ ($\chi=0.01$ and $\chi=0.1$) at Ra$_o=10^7$. In the previous models, these cases led to the emergence of a  subadiabatic region close to the outer boundary of the convection zone. Looking at the square of the non-dimensional buoyancy frequency profile  for this model (Fig. \ref{fig_18}), we observe that a slightly subadiabatic region does indeed appear, this time close to the inner boundary of the convection zone.
Note that the general mechanism for the appearance of this layer is the same as before, although the quantitative details differ. In this model setup, the mean kinetic energy is  again controlled only by the bulk Rayleigh number (see Fig. \ref{fig_6}), hence the velocity fluctuations remain large. However because $\beta(r)$ varies  between $0$ and $1$ the typical amplitude of the temperature perturbations $\Theta_{rms}$ is  much smaller (i.e. this time $\Theta_{rms}\propto \chi$). This results in a much smaller total  turbulent temperature flux  $\bar{F}_h$ compared with Models A and B.  As shown in Equation (\ref{N2eq}), however, whether a subadiabatic layer appears or not depends on the \textit{relative} amplitude of $\bar{F}_h$  compared to $\beta(r)$. Since $\beta(r)$ is close to $0$ near the inner  boundary, a small turbulent temperature flux is indeed sufficient to create a subadiabatic layer there {\color{black} according to Eq. (\ref{fluxeq})}.  
\par
\begin{figure}[H]
\centering
\includegraphics[scale=0.35]{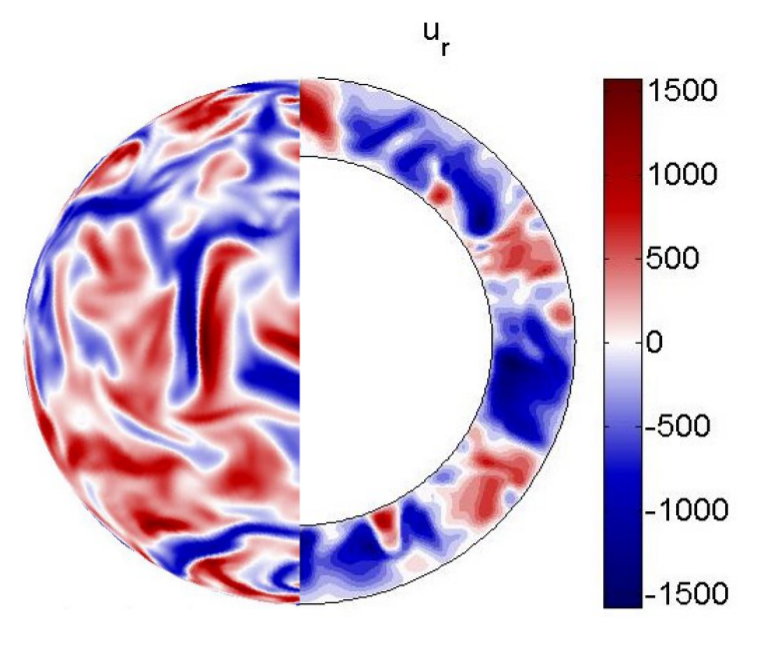}
\caption{\label{fig_19}
 (Color online) Snapshot of the radial velocity $u_r$ for $\chi=0.01$ and Ra$_o=10^7$ for Model C. The left part shows the $u_r$ field close to the outer radius just below the boundary layer. The right part shows the same field $u_r$ on a selected meridional plane.}
\end{figure}

 Figure \ref{fig_19} shows a snapshot  of $u_r$  for the $\chi=0.01$ case, in which the subadiabatic region is the deepest observed so far. Notice that convection is still vigorous throughout, again supporting our conclusions from previous models that this type of convection is highly non-local.

\section{Discussion}

\par We have studied  convection in a weakly compressible gaseous spherical shell, assuming a constant adiabatic temperature gradient as well as mixed temperature boundary conditions (fixed flux at the inner boundary and fixed temperature at the outer boundary). In Sections II, III and V, we presented results from three different model setups, that all have the same remarkable properties for sufficiently large Rayleigh number Ra, and sufficiently large variations in the superadiabaticity across the shell (measured by $\chi$). All these simulations showed substantial asymmetry between upflows and downflows, as well as the emergence of a subadiabatic layer which is still fully mixed by non-local convection. In Section IV, we explained these findings as follows{\color{black}:}
\par \textit{Asymmetry between upflows and downflows}: As in standard convection downward-traveling parcels are heated up by adiabatic compression, but remain cooler than the background temperature and therefore proceed to sink. In BRBC, these parcels would eventually have to warm up to match the temperature at the lower boundary, but in the case of fixed flux inner boundary condition, this is not the case and the parcels \textit{remain} cooler than the surroundings as they reach the bottom of the convection zone. There they {\color{black}rebound from} the boundary {\color{black}essentially being diverted (i.e. pressure-driven) into cool upflows}. We have found that for  large enough {\color{black} Ra$_o$}, all upflows are {\color{black}pressure-driven}. This asymmetric driving mechanism of upflows and downflows persists in much of the domain and proves that convection is highly non-local. {\color{black} These results are similar, though not identical, to those recently reported in \cite{Kapyla}}.
\par\textit{Emergence of a subadiabatic layer}: The fixed flux boundary condition at the inner boundary has a second consequence, namely that of tying the turbulent temperature flux to the perturbed diffusive temperature flux. Hence, for sufficiently large turbulent temperature flux, the diffusive temperature flux must also become large and can cause the background temperature gradient to exceed the adiabatic one{\color{black},  which results in a subadiabatic stratification.}
\par Two natural questions hence arise: What are the minimal necessary conditions for these dynamics to manifest themselves and why have these never been reported before {\color{black} in other Boussinesq studies}? As an answer  to the first question, we argue that the necessary conditions are (1) mixed temperature boundary conditions with fixed flux at one boundary and fixed temperature at the other{\color{black}, and (2)  sufficiently turbulent flows.} The importance of (1) should be clear from the description above. {\color{black} Condition (2)  is necessary for the turbulent fluxes to be large enough and for inertia to be strong enough.} These conditions are necessary, but do not have to be met  necessarily in exactly the same way {\color{black}they are created} as in Models A, B or C. For instance, we believe that with a sufficiently deep shell, it may be possible to achieve this dynamical regime even if $\chi$ were closer to one. {\color{black} Although not a strictly necessary condition, we have also found that having a radially-varying  superadiabaticity $\beta(r)$ more easily creates a large enough contrast across the domain and therefore lowers the threshold in Ra for the emergence of the subadiabatic layer. As such, it might even be possible to be in this unusual regime in a strictly Boussinesq Cartesian RBC, though the Rayleigh number may need to be extremely high, or one may need to invoke additional non-Boussinesq effects to generate a rapidly-varying $\beta$ and the therefore significant $\beta$ contrast (such as   varying gravity, for example, which causes a varying $dT_{\rm{ad}}/dz$, or by varying the diffusivities, or by adding internal heat sources within the fluid)}. By this reasoning, we {\color{black} conjecture that}  these dynamics may be found in high Pr number convection for large enough Ra.
\par Given these necessary conditions, we can now easily answer the second question. This kind of convection has not been previously observed {\color{black} in other Boussinesq studies} because the vast majority of investigations to date have used fixed temperature boundary conditions or fixed flux at both boundaries (e.g. \cite{Otero2002,Johnston}). There are certain studies in which  mixed temperature boundary conditions have been implemented, notably in  \cite{G78temp}. There,   low and intermediate Ra  were investigated, with a Prandtl number  equal to one and the flows were likely insufficiently turbulent (see condition {\color{black}2}) for the {\color{black}subadiabatic} layer to appear. In \cite{VZ08} (see also  \cite{SLZ11}),  turbulent convection in the high Ra  regime   using fixed flux at the bottom and fixed temperature at the top was also studied but there was no report of any subadiabatic  layer. Interestingly however, they indeed found larger rms temperatures near the lower boundary, similar to our results. They also noticed that the plumes were less buoyant and cooler and as a result  carried less heat compared with  cases where the temperature was fixed at both boundaries. In both cases however, the fluid was incompressible  with $dT_{\rm{ad}}/dz=0$ and $dT_{\rm{rad}}/dz$ constant, so that $\beta(z)=-1$ and there was much less imposed asymmetry in their {\color{black}system. Hence,}  although some prior studies have considered the effects of mixed temperature boundary conditions, ours appears to be the first to report the emergence of a subadiabatic layer. That implies that  a combination of {\color{black} both conditions described above} has to be satisfied. 
\par Finally, it is important to recall that we have used the SVB approximation for weakly compressible gases even though our model setups do not necessarily satisfy all the requirements of this approximation. Indeed, the two fundamental assumptions entering the SVB approximation are:
{\color{black} 1) that any perturbations of the thermodynamic quantities $\rho$ and $T$ about their domain mean $\rho_m$ and $T_m$ should be small, and 2) that the flow velocities be much smaller than the sound speed. These approximations then justify the use of equations (5) to (7). Note that SV also used a Cartesian domain, and further assumed, for simplicity, that $\alpha$, $g$, $\kappa$ and $\nu$, as well as the radiative and adiabatic temperature gradients were constant, but these are not  necessary conditions for the applicability of their equations. However, in a spherical geometry, or if these quantities vary with depth (as in our own models), one should verify that both the background state and the perturbations continue to satisfy conditions (1) and (2) a posteriori.  In Model A, as discussed in Section II-A, the sphericity of the domain implies that $T_{\rm{rad}}$ must vary with depth, even if everything else is held constant. As such, the SVB approximation can only be used if $\Delta T \ll T_m$, or equivalently, if $F_{\rm{rad}}(r_i^2/\kappa)(\frac{1}{r_i}-\frac{1}{r_o})\ll T_m$. To satisfy this condition and maintain a large Rayleigh number at the same time can be achieved by letting $\nu \rightarrow 0$ for instance. In Model B on the other hand, the validity of the SVB approximation is definitely violated, as $\alpha(r)$ varies quite significantly across the domain when $\chi$ is small. This invalidates the linearization of the equation of state used. A similar statement applies to Model C.}
 \par {\color{black}One} may therefore rightfully question whether the new dynamics discovered here are artifacts associated with breaking the bounds of validity of the SVB equations, or whether they would indeed  occur in a more realistic, fully compressible model setup as well. {\color{black}Based on published work and our own unpublished recent findings, the latter statement is likely true.  Indeed,  \cite{Chanetal}  reported the emergence of a subadiabatic layer close to the outer boundary of 3D, large-eddy simulations of compressible convection in an effectively plane parallel domain. However the resolution used was very low, shedding doubt at the time on the robustness of their results. Other studies have also noted the appearance of a subadiabatic mixed region in compressible hydrodynamic simulations (e.g. \cite{Mocak,Herwig}) but the setup in these cases was not based purely on a convection zone, but rather, on a convection zone embedded in between two stable regions. Recently,  \cite{Kapyla} also reported a subadiabatic layer in their fully compressible 3D simulations of overshooting convection. Nevertheless, a complete and definitive answer to this question requires the solution of the fully compressible equations, which is the subject of a future publication. Preliminary results obtained by the authors} with 3D fully compressible DNSs in a Cartesian box with mixed temperature boundary conditions do indeed show the appearance of a subadiabatic region, suggesting that the most salient feature of this kind of convection is robust {\color{black} (\cite{Brummelletal}, in prep.)}. We, however, expect the details to differ substantially, since the SVB approximation inherently suppresses some essential compressible dynamics, in particular the dynamic role of pressure in compressional heating.

\section*{ACKNOWLEDGMENTS}
The authors would like to thank C\'{e}line Guervilly for her help with the PARODY code {\color{black} and Hugues Faller for his useful comment}. This work was financially supported by NASA NNX14AG08G. The numerical simulations were performed on the Hyades supercomputer at UCSC purchased
using an NSF MRI grant.

\bibliographystyle{apsrev4-1}

\providecommand{\noopsort}[1]{}\providecommand{\singleletter}[1]{#1}%
\begin{thebibliography}{67}%
\makeatletter
\providecommand \@ifxundefined [1]{%
 \@ifx{#1\undefined}
}%
\providecommand \@ifnum [1]{%
 \ifnum #1\expandafter \@firstoftwo
 \else \expandafter \@secondoftwo
 \fi
}%
\providecommand \@ifx [1]{%
 \ifx #1\expandafter \@firstoftwo
 \else \expandafter \@secondoftwo
 \fi
}%
\providecommand \natexlab [1]{#1}%
\providecommand \enquote  [1]{``#1''}%
\providecommand \bibnamefont  [1]{#1}%
\providecommand \bibfnamefont [1]{#1}%
\providecommand \citenamefont [1]{#1}%
\providecommand \href@noop [0]{\@secondoftwo}%
\providecommand \href [0]{\begingroup \@sanitize@url \@href}%
\providecommand \@href[1]{\@@startlink{#1}\@@href}%
\providecommand \@@href[1]{\endgroup#1\@@endlink}%
\providecommand \@sanitize@url [0]{\catcode `\\12\catcode `\$12\catcode
  `\&12\catcode `\#12\catcode `\^12\catcode `\_12\catcode `\%12\relax}%
\providecommand \@@startlink[1]{}%
\providecommand \@@endlink[0]{}%
\providecommand \url  [0]{\begingroup\@sanitize@url \@url }%
\providecommand \@url [1]{\endgroup\@href {#1}{\urlprefix }}%
\providecommand \urlprefix  [0]{URL }%
\providecommand \Eprint [0]{\href }%
\providecommand \doibase [0]{http://dx.doi.org/}%
\providecommand \selectlanguage [0]{\@gobble}%
\providecommand \bibinfo  [0]{\@secondoftwo}%
\providecommand \bibfield  [0]{\@secondoftwo}%
\providecommand \translation [1]{[#1]}%
\providecommand \BibitemOpen [0]{}%
\providecommand \bibitemStop [0]{}%
\providecommand \bibitemNoStop [0]{.\EOS\space}%
\providecommand \EOS [0]{\spacefactor3000\relax}%
\providecommand \BibitemShut  [1]{\csname bibitem#1\endcsname}%
\let\auto@bib@innerbib\@empty
\bibitem [{\citenamefont {{Spiegel}}\ and\ \citenamefont
  {{Veronis}}(1960)}]{SV}%
  \BibitemOpen
  \bibfield  {author} {\bibinfo {author} {\bibfnamefont {E.~A.}\ \bibnamefont
  {{Spiegel}}}\ and\ \bibinfo {author} {\bibfnamefont {G.}~\bibnamefont
  {{Veronis}}},\ }\href@noop {} {\bibfield  {journal} {\bibinfo  {journal}
  {\apj}\ }\textbf {\bibinfo {volume} {131}},\ \bibinfo {pages} {442} (\bibinfo
  {year} {1960})}\BibitemShut {NoStop}%
\bibitem [{\citenamefont {{Bercovici}}\ \emph
  {et~al.}(1989{\natexlab{a}})\citenamefont {{Bercovici}}, \citenamefont
  {{Schubert}},\ and\ \citenamefont {{Glatzmaier}}}]{BSG89}%
  \BibitemOpen
  \bibfield  {author} {\bibinfo {author} {\bibfnamefont {D.}~\bibnamefont
  {{Bercovici}}}, \bibinfo {author} {\bibfnamefont {G.}~\bibnamefont
  {{Schubert}}}, \ and\ \bibinfo {author} {\bibfnamefont {G.~A.}\ \bibnamefont
  {{Glatzmaier}}},\ }\href@noop {} {\bibfield  {journal} {\bibinfo  {journal}
  {Science}\ }\textbf {\bibinfo {volume} {244}},\ \bibinfo {pages} {950}
  (\bibinfo {year} {1989}{\natexlab{a}})}\BibitemShut {NoStop}%
\bibitem [{\citenamefont {{Glatzmaier}}\ and\ \citenamefont
  {{Schubert}}(1993)}]{GS93}%
  \BibitemOpen
  \bibfield  {author} {\bibinfo {author} {\bibfnamefont {G.~A.}\ \bibnamefont
  {{Glatzmaier}}}\ and\ \bibinfo {author} {\bibfnamefont {G.}~\bibnamefont
  {{Schubert}}},\ }\href@noop {} {\bibfield  {journal} {\bibinfo  {journal}
  {Journal of Geophysical Research}\ }\textbf {\bibinfo {volume} {98}},\
  \bibinfo {pages} {21969} (\bibinfo {year} {1993})}\BibitemShut {NoStop}%
\bibitem [{\citenamefont {{Breuer}}\ \emph {et~al.}(2013)\citenamefont
  {{Breuer}}, \citenamefont {{Futterer}}, \citenamefont {{Plesa}},
  \citenamefont {{Krebs}}, \citenamefont {{Zaussinger}},\ and\ \citenamefont
  {{Egbers}}}]{exp}%
  \BibitemOpen
  \bibfield  {author} {\bibinfo {author} {\bibfnamefont {D.}~\bibnamefont
  {{Breuer}}}, \bibinfo {author} {\bibfnamefont {B.}~\bibnamefont
  {{Futterer}}}, \bibinfo {author} {\bibfnamefont {A.}~\bibnamefont {{Plesa}}},
  \bibinfo {author} {\bibfnamefont {A.}~\bibnamefont {{Krebs}}}, \bibinfo
  {author} {\bibfnamefont {F.}~\bibnamefont {{Zaussinger}}}, \ and\ \bibinfo
  {author} {\bibfnamefont {C.}~\bibnamefont {{Egbers}}},\ }\href@noop {}
  {\bibfield  {journal} {\bibinfo  {journal} {AGU Fall Meeting Abstracts}\ }
  (\bibinfo {year} {2013})}\BibitemShut {NoStop}%
\bibitem [{\citenamefont {{Schubert}}(1992)}]{mantle}%
  \BibitemOpen
  \bibfield  {author} {\bibinfo {author} {\bibfnamefont {G.}~\bibnamefont
  {{Schubert}}},\ }\href@noop {} {\bibfield  {journal} {\bibinfo  {journal}
  {Annual Review of Fluid Mechanics}\ }\textbf {\bibinfo {volume} {24}},\
  \bibinfo {pages} {359} (\bibinfo {year} {1992})}\BibitemShut {NoStop}%
\bibitem [{\citenamefont {{Marshall}}\ and\ \citenamefont
  {{Schott}}(1999)}]{ocean}%
  \BibitemOpen
  \bibfield  {author} {\bibinfo {author} {\bibfnamefont {J.}~\bibnamefont
  {{Marshall}}}\ and\ \bibinfo {author} {\bibfnamefont {F.}~\bibnamefont
  {{Schott}}},\ }\href@noop {} {\bibfield  {journal} {\bibinfo  {journal}
  {Reviews of Geophysics}\ }\textbf {\bibinfo {volume} {37}},\ \bibinfo {pages}
  {1} (\bibinfo {year} {1999})}\BibitemShut {NoStop}%
\bibitem [{\citenamefont {Emanuel}(1994)}]{atmo}%
  \BibitemOpen
  \bibfield  {author} {\bibinfo {author} {\bibfnamefont {K.~A.}\ \bibnamefont
  {Emanuel}},\ }\href@noop {} {\emph {\bibinfo {title} {Atmospheric
  convection}}}\ (\bibinfo  {publisher} {Oxford University Press},\ \bibinfo
  {year} {1994})\BibitemShut {NoStop}%
\bibitem [{\citenamefont {{Rieutord}}\ and\ \citenamefont
  {{Rincon}}(2010)}]{RR10}%
  \BibitemOpen
  \bibfield  {author} {\bibinfo {author} {\bibfnamefont {M.}~\bibnamefont
  {{Rieutord}}}\ and\ \bibinfo {author} {\bibfnamefont {F.}~\bibnamefont
  {{Rincon}}},\ }\href@noop {} {\bibfield  {journal} {\bibinfo  {journal}
  {Living Reviews in Solar Physics}\ }\textbf {\bibinfo {volume} {7}},\
  \bibinfo {eid} {2} (\bibinfo {year} {2010})}\BibitemShut {NoStop}%
\bibitem [{\citenamefont {{Gough}}\ \emph {et~al.}(1996)\citenamefont
  {{Gough}}, \citenamefont {{Kosovichev}}, \citenamefont {{Toomre}},
  \citenamefont {{Anderson}}, \citenamefont {{Antia}}, \citenamefont {{Basu}},
  \citenamefont {{Chaboyer}}, \citenamefont {{Chitre}}, \citenamefont
  {{Christensen-Dalsgaard}}, \citenamefont {{Dziembowski}}, \citenamefont
  {{Eff-Darwich}}, \citenamefont {{Elliott}}, \citenamefont {{Giles}},
  \citenamefont {{Goode}}, \citenamefont {{Guzik}}, \citenamefont {{Harvey}},
  \citenamefont {{Hill}}, \citenamefont {{Leibacher}}, \citenamefont
  {{Monteiro}}, \citenamefont {{Richard}}, \citenamefont {{Sekii}},
  \citenamefont {{Shibahashi}}, \citenamefont {{Takata}}, \citenamefont
  {{Thompson}}, \citenamefont {{Vauclair}},\ and\ \citenamefont
  {{Vorontsov}}}]{astro}%
  \BibitemOpen
  \bibfield  {author} {\bibinfo {author} {\bibfnamefont {D.~O.}\ \bibnamefont
  {{Gough}}}, \bibinfo {author} {\bibfnamefont {A.~G.}\ \bibnamefont
  {{Kosovichev}}}, \bibinfo {author} {\bibfnamefont {J.}~\bibnamefont
  {{Toomre}}}, \bibinfo {author} {\bibfnamefont {E.}~\bibnamefont
  {{Anderson}}}, \bibinfo {author} {\bibfnamefont {H.~M.}\ \bibnamefont
  {{Antia}}}, \bibinfo {author} {\bibfnamefont {S.}~\bibnamefont {{Basu}}},
  \bibinfo {author} {\bibfnamefont {B.}~\bibnamefont {{Chaboyer}}}, \bibinfo
  {author} {\bibfnamefont {S.~M.}\ \bibnamefont {{Chitre}}}, \bibinfo {author}
  {\bibfnamefont {J.}~\bibnamefont {{Christensen-Dalsgaard}}}, \bibinfo
  {author} {\bibfnamefont {W.~A.}\ \bibnamefont {{Dziembowski}}}, \bibinfo
  {author} {\bibfnamefont {A.}~\bibnamefont {{Eff-Darwich}}}, \bibinfo {author}
  {\bibfnamefont {J.~R.}\ \bibnamefont {{Elliott}}}, \bibinfo {author}
  {\bibfnamefont {P.~M.}\ \bibnamefont {{Giles}}}, \bibinfo {author}
  {\bibfnamefont {P.~R.}\ \bibnamefont {{Goode}}}, \bibinfo {author}
  {\bibfnamefont {J.~A.}\ \bibnamefont {{Guzik}}}, \bibinfo {author}
  {\bibfnamefont {J.~W.}\ \bibnamefont {{Harvey}}}, \bibinfo {author}
  {\bibfnamefont {F.}~\bibnamefont {{Hill}}}, \bibinfo {author} {\bibfnamefont
  {J.~W.}\ \bibnamefont {{Leibacher}}}, \bibinfo {author} {\bibfnamefont
  {M.~J.~P.~F.~G.}\ \bibnamefont {{Monteiro}}}, \bibinfo {author}
  {\bibfnamefont {O.}~\bibnamefont {{Richard}}}, \bibinfo {author}
  {\bibfnamefont {T.}~\bibnamefont {{Sekii}}}, \bibinfo {author} {\bibfnamefont
  {H.}~\bibnamefont {{Shibahashi}}}, \bibinfo {author} {\bibfnamefont
  {M.}~\bibnamefont {{Takata}}}, \bibinfo {author} {\bibfnamefont {M.~J.}\
  \bibnamefont {{Thompson}}}, \bibinfo {author} {\bibfnamefont
  {S.}~\bibnamefont {{Vauclair}}}, \ and\ \bibinfo {author} {\bibfnamefont
  {S.~V.}\ \bibnamefont {{Vorontsov}}},\ }\href@noop {} {\bibfield  {journal}
  {\bibinfo  {journal} {Science}\ }\textbf {\bibinfo {volume} {272}},\ \bibinfo
  {pages} {1296} (\bibinfo {year} {1996})}\BibitemShut {NoStop}%
\bibitem [{\citenamefont {{Boussinesq}}(1903)}]{Bouss}%
  \BibitemOpen
  \bibfield  {author} {\bibinfo {author} {\bibfnamefont {J.}~\bibnamefont
  {{Boussinesq}}},\ }\href@noop {} {\emph {\bibinfo {title} {{Theorie
  Analytique de la Chaleur}}}},\ Vol.~\bibinfo {volume} {2}\ (\bibinfo
  {publisher} {Gauthier Villars, Paris},\ \bibinfo {year} {1903})\BibitemShut
  {NoStop}%
\bibitem [{\citenamefont {{Oberbeck}}(1879)}]{Ober}%
  \BibitemOpen
  \bibfield  {author} {\bibinfo {author} {\bibfnamefont {A.}~\bibnamefont
  {{Oberbeck}}},\ }\href@noop {} {\bibfield  {journal} {\bibinfo  {journal}
  {Annalen der Physik}\ }\textbf {\bibinfo {volume} {243}},\ \bibinfo {pages}
  {271} (\bibinfo {year} {1879})}\BibitemShut {NoStop}%
\bibitem [{\citenamefont {{Veronis}}(1966)}]{Veronis66}%
  \BibitemOpen
  \bibfield  {author} {\bibinfo {author} {\bibfnamefont {G.}~\bibnamefont
  {{Veronis}}},\ }\href@noop {} {\bibfield  {journal} {\bibinfo  {journal}
  {Journal of Fluid Mechanics}\ }\textbf {\bibinfo {volume} {26}},\ \bibinfo
  {pages} {49} (\bibinfo {year} {1966})}\BibitemShut {NoStop}%
\bibitem [{\citenamefont {{Moore}}\ and\ \citenamefont {{Weiss}}(1973)}]{MW73}%
  \BibitemOpen
  \bibfield  {author} {\bibinfo {author} {\bibfnamefont {D.~R.}\ \bibnamefont
  {{Moore}}}\ and\ \bibinfo {author} {\bibfnamefont {N.~O.}\ \bibnamefont
  {{Weiss}}},\ }\href@noop {} {\bibfield  {journal} {\bibinfo  {journal}
  {Journal of Fluid Mechanics}\ }\textbf {\bibinfo {volume} {58}},\ \bibinfo
  {pages} {289} (\bibinfo {year} {1973})}\BibitemShut {NoStop}%
\bibitem [{\citenamefont {{Shraiman}}\ and\ \citenamefont
  {{Siggia}}(1990)}]{SS90}%
  \BibitemOpen
  \bibfield  {author} {\bibinfo {author} {\bibfnamefont {B.~I.}\ \bibnamefont
  {{Shraiman}}}\ and\ \bibinfo {author} {\bibfnamefont {E.~D.}\ \bibnamefont
  {{Siggia}}},\ }\href@noop {} {\bibfield  {journal} {\bibinfo  {journal}
  {\pra}\ }\textbf {\bibinfo {volume} {42}},\ \bibinfo {pages} {3650} (\bibinfo
  {year} {1990})}\BibitemShut {NoStop}%
\bibitem [{\citenamefont {{Julien}}\ \emph {et~al.}(1996)\citenamefont
  {{Julien}}, \citenamefont {{Legg}}, \citenamefont {{McWilliams}},\ and\
  \citenamefont {{Werne}}}]{Julien}%
  \BibitemOpen
  \bibfield  {author} {\bibinfo {author} {\bibfnamefont {K.}~\bibnamefont
  {{Julien}}}, \bibinfo {author} {\bibfnamefont {S.}~\bibnamefont {{Legg}}},
  \bibinfo {author} {\bibfnamefont {J.}~\bibnamefont {{McWilliams}}}, \ and\
  \bibinfo {author} {\bibfnamefont {J.}~\bibnamefont {{Werne}}},\ }\href@noop
  {} {\bibfield  {journal} {\bibinfo  {journal} {Journal of Fluid Mechanics}\
  }\textbf {\bibinfo {volume} {322}},\ \bibinfo {pages} {243} (\bibinfo {year}
  {1996})}\BibitemShut {NoStop}%
\bibitem [{\citenamefont {{Guervilly}}\ \emph {et~al.}(2014)\citenamefont
  {{Guervilly}}, \citenamefont {{Hughes}},\ and\ \citenamefont
  {{Jones}}}]{Celine}%
  \BibitemOpen
  \bibfield  {author} {\bibinfo {author} {\bibfnamefont {C.}~\bibnamefont
  {{Guervilly}}}, \bibinfo {author} {\bibfnamefont {D.~W.}\ \bibnamefont
  {{Hughes}}}, \ and\ \bibinfo {author} {\bibfnamefont {C.~A.}\ \bibnamefont
  {{Jones}}},\ }\href@noop {} {\bibfield  {journal} {\bibinfo  {journal}
  {Journal of Fluid Mechanics}\ }\textbf {\bibinfo {volume} {758}},\ \bibinfo
  {pages} {407} (\bibinfo {year} {2014})}\BibitemShut {NoStop}%
\bibitem [{\citenamefont {{Ahlers}}\ \emph {et~al.}(2009)\citenamefont
  {{Ahlers}}, \citenamefont {{Grossmann}},\ and\ \citenamefont
  {{Lohse}}}]{review}%
  \BibitemOpen
  \bibfield  {author} {\bibinfo {author} {\bibfnamefont {G.}~\bibnamefont
  {{Ahlers}}}, \bibinfo {author} {\bibfnamefont {S.}~\bibnamefont
  {{Grossmann}}}, \ and\ \bibinfo {author} {\bibfnamefont {D.}~\bibnamefont
  {{Lohse}}},\ }\href@noop {} {\bibfield  {journal} {\bibinfo  {journal}
  {Reviews of Modern Physics}\ }\textbf {\bibinfo {volume} {81}},\ \bibinfo
  {pages} {503} (\bibinfo {year} {2009})}\BibitemShut {NoStop}%
\bibitem [{\citenamefont {Chandrasekhar}(1961)}]{chandra}%
  \BibitemOpen
  \bibfield  {author} {\bibinfo {author} {\bibfnamefont {S.}~\bibnamefont
  {Chandrasekhar}},\ }\href@noop {} {\emph {\bibinfo {title} {Hydrodynamic and
  Hydromagnetic Stability}}}\ (\bibinfo  {publisher} {Oxford University
  Press},\ \bibinfo {year} {1961})\BibitemShut {NoStop}%
\bibitem [{\citenamefont {{Busse}}(1975)}]{Busse}%
  \BibitemOpen
  \bibfield  {author} {\bibinfo {author} {\bibfnamefont {F.~H.}\ \bibnamefont
  {{Busse}}},\ }\href@noop {} {\bibfield  {journal} {\bibinfo  {journal}
  {Journal of Fluid Mechanics}\ }\textbf {\bibinfo {volume} {72}},\ \bibinfo
  {pages} {67} (\bibinfo {year} {1975})}\BibitemShut {NoStop}%
\bibitem [{\citenamefont {{Zebib}}\ \emph {et~al.}(1980)\citenamefont
  {{Zebib}}, \citenamefont {{Schubert}},\ and\ \citenamefont
  {{Straus}}}]{Zebib80}%
  \BibitemOpen
  \bibfield  {author} {\bibinfo {author} {\bibfnamefont {A.}~\bibnamefont
  {{Zebib}}}, \bibinfo {author} {\bibfnamefont {G.}~\bibnamefont {{Schubert}}},
  \ and\ \bibinfo {author} {\bibfnamefont {J.~M.}\ \bibnamefont {{Straus}}},\
  }\href@noop {} {\bibfield  {journal} {\bibinfo  {journal} {Journal of Fluid
  Mechanics}\ }\textbf {\bibinfo {volume} {97}},\ \bibinfo {pages} {257}
  (\bibinfo {year} {1980})}\BibitemShut {NoStop}%
\bibitem [{\citenamefont {{Schubert}}\ and\ \citenamefont
  {{Zebib}}(1980)}]{SZ80}%
  \BibitemOpen
  \bibfield  {author} {\bibinfo {author} {\bibfnamefont {G.}~\bibnamefont
  {{Schubert}}}\ and\ \bibinfo {author} {\bibfnamefont {A.}~\bibnamefont
  {{Zebib}}},\ }\href@noop {} {\bibfield  {journal} {\bibinfo  {journal}
  {Geophysical and Astrophysical Fluid Dynamics}\ }\textbf {\bibinfo {volume}
  {15}},\ \bibinfo {pages} {65} (\bibinfo {year} {1980})}\BibitemShut {NoStop}%
\bibitem [{\citenamefont {{Zebib}}\ \emph {et~al.}(1985)\citenamefont
  {{Zebib}}, \citenamefont {{Goyal}},\ and\ \citenamefont
  {{Schubert}}}]{Zebib85}%
  \BibitemOpen
  \bibfield  {author} {\bibinfo {author} {\bibfnamefont {A.}~\bibnamefont
  {{Zebib}}}, \bibinfo {author} {\bibfnamefont {A.~K.}\ \bibnamefont
  {{Goyal}}}, \ and\ \bibinfo {author} {\bibfnamefont {G.}~\bibnamefont
  {{Schubert}}},\ }\href@noop {} {\bibfield  {journal} {\bibinfo  {journal}
  {Journal of Fluid Mechanics}\ }\textbf {\bibinfo {volume} {152}},\ \bibinfo
  {pages} {39} (\bibinfo {year} {1985})}\BibitemShut {NoStop}%
\bibitem [{\citenamefont {{Machetel}}\ \emph {et~al.}(1986)\citenamefont
  {{Machetel}}, \citenamefont {{Rabinowicz}},\ and\ \citenamefont
  {{Bernardet}}}]{Machetel}%
  \BibitemOpen
  \bibfield  {author} {\bibinfo {author} {\bibfnamefont {P.}~\bibnamefont
  {{Machetel}}}, \bibinfo {author} {\bibfnamefont {M.}~\bibnamefont
  {{Rabinowicz}}}, \ and\ \bibinfo {author} {\bibfnamefont {P.}~\bibnamefont
  {{Bernardet}}},\ }\href@noop {} {\bibfield  {journal} {\bibinfo  {journal}
  {Geophysical and Astrophysical Fluid Dynamics}\ }\textbf {\bibinfo {volume}
  {37}},\ \bibinfo {pages} {57} (\bibinfo {year} {1986})}\BibitemShut {NoStop}%
\bibitem [{\citenamefont {{Bercovici}}\ \emph
  {et~al.}(1989{\natexlab{b}})\citenamefont {{Bercovici}}, \citenamefont
  {{Schubert}}, \citenamefont {{Glatzmaier}},\ and\ \citenamefont
  {{Zebib}}}]{Bercovici}%
  \BibitemOpen
  \bibfield  {author} {\bibinfo {author} {\bibfnamefont {D.}~\bibnamefont
  {{Bercovici}}}, \bibinfo {author} {\bibfnamefont {G.}~\bibnamefont
  {{Schubert}}}, \bibinfo {author} {\bibfnamefont {G.~A.}\ \bibnamefont
  {{Glatzmaier}}}, \ and\ \bibinfo {author} {\bibfnamefont {A.}~\bibnamefont
  {{Zebib}}},\ }\href@noop {} {\bibfield  {journal} {\bibinfo  {journal}
  {Journal of Fluid Mechanics}\ }\textbf {\bibinfo {volume} {206}},\ \bibinfo
  {pages} {75} (\bibinfo {year} {1989}{\natexlab{b}})}\BibitemShut {NoStop}%
\bibitem [{\citenamefont {{Vangelov}}\ and\ \citenamefont
  {{Jarvis}}(1994)}]{Vangelov}%
  \BibitemOpen
  \bibfield  {author} {\bibinfo {author} {\bibfnamefont {V.~I.}\ \bibnamefont
  {{Vangelov}}}\ and\ \bibinfo {author} {\bibfnamefont {G.~T.}\ \bibnamefont
  {{Jarvis}}},\ }\href@noop {} {\bibfield  {journal} {\bibinfo  {journal}
  {Journal of Geophysical Research}\ }\textbf {\bibinfo {volume} {99}},\
  \bibinfo {pages} {9345} (\bibinfo {year} {1994})}\BibitemShut {NoStop}%
\bibitem [{\citenamefont {{Jarvis}}\ \emph {et~al.}(1995)\citenamefont
  {{Jarvis}}, \citenamefont {{Glatzmaier}},\ and\ \citenamefont
  {{Vangelov}}}]{Jarvis}%
  \BibitemOpen
  \bibfield  {author} {\bibinfo {author} {\bibfnamefont {G.~T.}\ \bibnamefont
  {{Jarvis}}}, \bibinfo {author} {\bibfnamefont {G.~A.}\ \bibnamefont
  {{Glatzmaier}}}, \ and\ \bibinfo {author} {\bibfnamefont {V.~I.}\
  \bibnamefont {{Vangelov}}},\ }\href@noop {} {\bibfield  {journal} {\bibinfo
  {journal} {Geophysical and Astrophysical Fluid Dynamics}\ }\textbf {\bibinfo
  {volume} {79}},\ \bibinfo {pages} {147} (\bibinfo {year} {1995})}\BibitemShut
  {NoStop}%
\bibitem [{\citenamefont {{Choblet}}(2012)}]{Choblet}%
  \BibitemOpen
  \bibfield  {author} {\bibinfo {author} {\bibfnamefont {G.}~\bibnamefont
  {{Choblet}}},\ }\href@noop {} {\bibfield  {journal} {\bibinfo  {journal}
  {Physics of the Earth and Planetary Interiors}\ }\textbf {\bibinfo {volume}
  {206}},\ \bibinfo {pages} {31} (\bibinfo {year} {2012})}\BibitemShut
  {NoStop}%
\bibitem [{\citenamefont {{Dormy}}\ \emph {et~al.}(2004)\citenamefont
  {{Dormy}}, \citenamefont {{Soward}}, \citenamefont {{Jones}}, \citenamefont
  {{Jault}},\ and\ \citenamefont {{Cardin}}}]{Dormy}%
  \BibitemOpen
  \bibfield  {author} {\bibinfo {author} {\bibfnamefont {E.}~\bibnamefont
  {{Dormy}}}, \bibinfo {author} {\bibfnamefont {A.~M.}\ \bibnamefont
  {{Soward}}}, \bibinfo {author} {\bibfnamefont {C.~A.}\ \bibnamefont
  {{Jones}}}, \bibinfo {author} {\bibfnamefont {D.}~\bibnamefont {{Jault}}}, \
  and\ \bibinfo {author} {\bibfnamefont {P.}~\bibnamefont {{Cardin}}},\
  }\href@noop {} {\bibfield  {journal} {\bibinfo  {journal} {Journal of Fluid
  Mechanics}\ }\textbf {\bibinfo {volume} {501}},\ \bibinfo {pages} {43}
  (\bibinfo {year} {2004})}\BibitemShut {NoStop}%
\bibitem [{\citenamefont {{Tilgner}}(1996)}]{Tilgner96}%
  \BibitemOpen
  \bibfield  {author} {\bibinfo {author} {\bibfnamefont {A.}~\bibnamefont
  {{Tilgner}}},\ }\href@noop {} {\bibfield  {journal} {\bibinfo  {journal}
  {\pre}\ }\textbf {\bibinfo {volume} {53}},\ \bibinfo {pages} {4847} (\bibinfo
  {year} {1996})}\BibitemShut {NoStop}%
\bibitem [{\citenamefont {{Feudel}}\ \emph {et~al.}(2011)\citenamefont
  {{Feudel}}, \citenamefont {{Bergemann}}, \citenamefont {{Tuckerman}},
  \citenamefont {{Egbers}}, \citenamefont {{Futterer}}, \citenamefont
  {{Gellert}},\ and\ \citenamefont {{Hollerbach}}}]{Feudel11}%
  \BibitemOpen
  \bibfield  {author} {\bibinfo {author} {\bibfnamefont {F.}~\bibnamefont
  {{Feudel}}}, \bibinfo {author} {\bibfnamefont {K.}~\bibnamefont
  {{Bergemann}}}, \bibinfo {author} {\bibfnamefont {L.~S.}\ \bibnamefont
  {{Tuckerman}}}, \bibinfo {author} {\bibfnamefont {C.}~\bibnamefont
  {{Egbers}}}, \bibinfo {author} {\bibfnamefont {B.}~\bibnamefont
  {{Futterer}}}, \bibinfo {author} {\bibfnamefont {M.}~\bibnamefont
  {{Gellert}}}, \ and\ \bibinfo {author} {\bibfnamefont {R.}~\bibnamefont
  {{Hollerbach}}},\ }\href@noop {} {\bibfield  {journal} {\bibinfo  {journal}
  {\pre}\ }\textbf {\bibinfo {volume} {83}},\ \bibinfo {eid} {046304} (\bibinfo
  {year} {2011})}\BibitemShut {NoStop}%
\bibitem [{\citenamefont {{Gastine}}\ \emph {et~al.}(2015)\citenamefont
  {{Gastine}}, \citenamefont {{Wicht}},\ and\ \citenamefont
  {{Aurnou}}}]{Gastine15}%
  \BibitemOpen
  \bibfield  {author} {\bibinfo {author} {\bibfnamefont {T.}~\bibnamefont
  {{Gastine}}}, \bibinfo {author} {\bibfnamefont {J.}~\bibnamefont {{Wicht}}},
  \ and\ \bibinfo {author} {\bibfnamefont {J.~M.}\ \bibnamefont {{Aurnou}}},\
  }\href@noop {} {\bibfield  {journal} {\bibinfo  {journal} {Journal of Fluid
  Mechanics}\ }\textbf {\bibinfo {volume} {778}},\ \bibinfo {pages} {721}
  (\bibinfo {year} {2015})}\BibitemShut {NoStop}%
\bibitem [{\citenamefont {{Gastine}}\ \emph {et~al.}(2016)\citenamefont
  {{Gastine}}, \citenamefont {{Wicht}},\ and\ \citenamefont
  {{Aubert}}}]{Gastine16}%
  \BibitemOpen
  \bibfield  {author} {\bibinfo {author} {\bibfnamefont {T.}~\bibnamefont
  {{Gastine}}}, \bibinfo {author} {\bibfnamefont {J.}~\bibnamefont {{Wicht}}},
  \ and\ \bibinfo {author} {\bibfnamefont {J.}~\bibnamefont {{Aubert}}},\
  }\href@noop {} {\bibfield  {journal} {\bibinfo  {journal} {Journal of Fluid
  Mechanics}\ }\textbf {\bibinfo {volume} {808}},\ \bibinfo {pages} {690}
  (\bibinfo {year} {2016})}\BibitemShut {NoStop}%
\bibitem [{\citenamefont {{Grossmann}}\ and\ \citenamefont
  {{Lohse}}(2000)}]{GL2000}%
  \BibitemOpen
  \bibfield  {author} {\bibinfo {author} {\bibfnamefont {S.}~\bibnamefont
  {{Grossmann}}}\ and\ \bibinfo {author} {\bibfnamefont {D.}~\bibnamefont
  {{Lohse}}},\ }\href@noop {} {\bibfield  {journal} {\bibinfo  {journal}
  {Journal of Fluid Mechanics}\ }\textbf {\bibinfo {volume} {407}},\ \bibinfo
  {pages} {27} (\bibinfo {year} {2000})}\BibitemShut {NoStop}%
\bibitem [{\citenamefont {{Batchelor}}(1953)}]{Batchelor}%
  \BibitemOpen
  \bibfield  {author} {\bibinfo {author} {\bibfnamefont {G.~K.}\ \bibnamefont
  {{Batchelor}}},\ }\href@noop {} {\bibfield  {journal} {\bibinfo  {journal}
  {Quarterly Journal of the Royal Meteorological Society}\ }\textbf {\bibinfo
  {volume} {79}},\ \bibinfo {pages} {224} (\bibinfo {year} {1953})}\BibitemShut
  {NoStop}%
\bibitem [{\citenamefont {{Ogura}}\ and\ \citenamefont
  {{Phillips}}(1962)}]{Ogura}%
  \BibitemOpen
  \bibfield  {author} {\bibinfo {author} {\bibfnamefont {Y.}~\bibnamefont
  {{Ogura}}}\ and\ \bibinfo {author} {\bibfnamefont {N.~A.}\ \bibnamefont
  {{Phillips}}},\ }\href@noop {} {\bibfield  {journal} {\bibinfo  {journal}
  {Journal of Atmospheric Sciences}\ }\textbf {\bibinfo {volume} {19}},\
  \bibinfo {pages} {173} (\bibinfo {year} {1962})}\BibitemShut {NoStop}%
\bibitem [{\citenamefont {{Gough}}(1969)}]{Gough69}%
  \BibitemOpen
  \bibfield  {author} {\bibinfo {author} {\bibfnamefont {D.~O.}\ \bibnamefont
  {{Gough}}},\ }\href@noop {} {\bibfield  {journal} {\bibinfo  {journal}
  {Journal of Atmospheric Sciences}\ }\textbf {\bibinfo {volume} {26}},\
  \bibinfo {pages} {448} (\bibinfo {year} {1969})}\BibitemShut {NoStop}%
\bibitem [{\citenamefont {{Latour}}\ \emph {et~al.}(1976)\citenamefont
  {{Latour}}, \citenamefont {{Spiegel}}, \citenamefont {{Toomre}},\ and\
  \citenamefont {{Zahn}}}]{Latour}%
  \BibitemOpen
  \bibfield  {author} {\bibinfo {author} {\bibfnamefont {J.}~\bibnamefont
  {{Latour}}}, \bibinfo {author} {\bibfnamefont {E.~A.}\ \bibnamefont
  {{Spiegel}}}, \bibinfo {author} {\bibfnamefont {J.}~\bibnamefont {{Toomre}}},
  \ and\ \bibinfo {author} {\bibfnamefont {J.-P.}\ \bibnamefont {{Zahn}}},\
  }\href@noop {} {\bibfield  {journal} {\bibinfo  {journal} {\apj}\ }\textbf
  {\bibinfo {volume} {207}},\ \bibinfo {pages} {233} (\bibinfo {year}
  {1976})}\BibitemShut {NoStop}%
\bibitem [{\citenamefont {{Gilman}}\ and\ \citenamefont
  {{Glatzmaier}}(1981)}]{GG}%
  \BibitemOpen
  \bibfield  {author} {\bibinfo {author} {\bibfnamefont {P.~A.}\ \bibnamefont
  {{Gilman}}}\ and\ \bibinfo {author} {\bibfnamefont {G.~A.}\ \bibnamefont
  {{Glatzmaier}}},\ }\href@noop {} {\bibfield  {journal} {\bibinfo  {journal}
  {Astrophysical Journal Supplement Series}\ }\textbf {\bibinfo {volume}
  {45}},\ \bibinfo {pages} {335} (\bibinfo {year} {1981})}\BibitemShut
  {NoStop}%
\bibitem [{\citenamefont {{Miesch}}(2005)}]{Miesch}%
  \BibitemOpen
  \bibfield  {author} {\bibinfo {author} {\bibfnamefont {M.~S.}\ \bibnamefont
  {{Miesch}}},\ }\href@noop {} {\bibfield  {journal} {\bibinfo  {journal}
  {Living Reviews in Solar Physics}\ }\textbf {\bibinfo {volume} {2}},\
  \bibinfo {eid} {1} (\bibinfo {year} {2005})}\BibitemShut {NoStop}%
\bibitem [{\citenamefont {{Browning}}\ \emph {et~al.}(2004)\citenamefont
  {{Browning}}, \citenamefont {{Brun}},\ and\ \citenamefont {{Toomre}}}]{BBT}%
  \BibitemOpen
  \bibfield  {author} {\bibinfo {author} {\bibfnamefont {M.~K.}\ \bibnamefont
  {{Browning}}}, \bibinfo {author} {\bibfnamefont {A.~S.}\ \bibnamefont
  {{Brun}}}, \ and\ \bibinfo {author} {\bibfnamefont {J.}~\bibnamefont
  {{Toomre}}},\ }\href@noop {} {\bibfield  {journal} {\bibinfo  {journal}
  {\apj}\ }\textbf {\bibinfo {volume} {601}},\ \bibinfo {pages} {512} (\bibinfo
  {year} {2004})}\BibitemShut {NoStop}%
\bibitem [{\citenamefont {{Brown}}\ \emph {et~al.}(2008)\citenamefont
  {{Brown}}, \citenamefont {{Browning}}, \citenamefont {{Brun}}, \citenamefont
  {{Miesch}},\ and\ \citenamefont {{Toomre}}}]{BBB}%
  \BibitemOpen
  \bibfield  {author} {\bibinfo {author} {\bibfnamefont {B.~P.}\ \bibnamefont
  {{Brown}}}, \bibinfo {author} {\bibfnamefont {M.~K.}\ \bibnamefont
  {{Browning}}}, \bibinfo {author} {\bibfnamefont {A.~S.}\ \bibnamefont
  {{Brun}}}, \bibinfo {author} {\bibfnamefont {M.~S.}\ \bibnamefont
  {{Miesch}}}, \ and\ \bibinfo {author} {\bibfnamefont {J.}~\bibnamefont
  {{Toomre}}},\ }\href@noop {} {\bibfield  {journal} {\bibinfo  {journal}
  {\apj}\ }\textbf {\bibinfo {volume} {689}},\ \bibinfo {eid} {1354-1372}
  (\bibinfo {year} {2008})}\BibitemShut {NoStop}%
\bibitem [{\citenamefont {{Augustson}}\ \emph {et~al.}(2012)\citenamefont
  {{Augustson}}, \citenamefont {{Brown}}, \citenamefont {{Brun}}, \citenamefont
  {{Miesch}},\ and\ \citenamefont {{Toomre}}}]{Aug12}%
  \BibitemOpen
  \bibfield  {author} {\bibinfo {author} {\bibfnamefont {K.~C.}\ \bibnamefont
  {{Augustson}}}, \bibinfo {author} {\bibfnamefont {B.~P.}\ \bibnamefont
  {{Brown}}}, \bibinfo {author} {\bibfnamefont {A.~S.}\ \bibnamefont {{Brun}}},
  \bibinfo {author} {\bibfnamefont {M.~S.}\ \bibnamefont {{Miesch}}}, \ and\
  \bibinfo {author} {\bibfnamefont {J.}~\bibnamefont {{Toomre}}},\ }\href@noop
  {} {\bibfield  {journal} {\bibinfo  {journal} {\apj}\ }\textbf {\bibinfo
  {volume} {756}},\ \bibinfo {eid} {169} (\bibinfo {year} {2012})}\BibitemShut
  {NoStop}%
\bibitem [{\citenamefont {{Brown}}\ \emph {et~al.}(2012)\citenamefont
  {{Brown}}, \citenamefont {{Vasil}},\ and\ \citenamefont {{Zweibel}}}]{BVZ}%
  \BibitemOpen
  \bibfield  {author} {\bibinfo {author} {\bibfnamefont {B.~P.}\ \bibnamefont
  {{Brown}}}, \bibinfo {author} {\bibfnamefont {G.~M.}\ \bibnamefont
  {{Vasil}}}, \ and\ \bibinfo {author} {\bibfnamefont {E.~G.}\ \bibnamefont
  {{Zweibel}}},\ }\href@noop {} {\bibfield  {journal} {\bibinfo  {journal}
  {\apj}\ }\textbf {\bibinfo {volume} {756}},\ \bibinfo {eid} {109} (\bibinfo
  {year} {2012})}\BibitemShut {NoStop}%
\bibitem [{\citenamefont {{Vasil}}\ \emph {et~al.}(2013)\citenamefont
  {{Vasil}}, \citenamefont {{Lecoanet}}, \citenamefont {{Brown}}, \citenamefont
  {{Wood}},\ and\ \citenamefont {{Zweibel}}}]{Vasil}%
  \BibitemOpen
  \bibfield  {author} {\bibinfo {author} {\bibfnamefont {G.~M.}\ \bibnamefont
  {{Vasil}}}, \bibinfo {author} {\bibfnamefont {D.}~\bibnamefont {{Lecoanet}}},
  \bibinfo {author} {\bibfnamefont {B.~P.}\ \bibnamefont {{Brown}}}, \bibinfo
  {author} {\bibfnamefont {T.~S.}\ \bibnamefont {{Wood}}}, \ and\ \bibinfo
  {author} {\bibfnamefont {E.~G.}\ \bibnamefont {{Zweibel}}},\ }\href@noop {}
  {\bibfield  {journal} {\bibinfo  {journal} {\apj}\ }\textbf {\bibinfo
  {volume} {773}},\ \bibinfo {eid} {169} (\bibinfo {year} {2013})}\BibitemShut
  {NoStop}%
\bibitem [{\citenamefont {{Verhoeven}}\ \emph {et~al.}(2015)\citenamefont
  {{Verhoeven}}, \citenamefont {{Wieseh{\"o}fer}},\ and\ \citenamefont
  {{Stellmach}}}]{ancom}%
  \BibitemOpen
  \bibfield  {author} {\bibinfo {author} {\bibfnamefont {J.}~\bibnamefont
  {{Verhoeven}}}, \bibinfo {author} {\bibfnamefont {T.}~\bibnamefont
  {{Wieseh{\"o}fer}}}, \ and\ \bibinfo {author} {\bibfnamefont
  {S.}~\bibnamefont {{Stellmach}}},\ }\href@noop {} {\bibfield  {journal}
  {\bibinfo  {journal} {\apj}\ }\textbf {\bibinfo {volume} {805}},\ \bibinfo
  {eid} {62} (\bibinfo {year} {2015})}\BibitemShut {NoStop}%
\bibitem [{\citenamefont {{Gilman}}(1976)}]{Gilman76}%
  \BibitemOpen
  \bibfield  {author} {\bibinfo {author} {\bibfnamefont {P.~A.}\ \bibnamefont
  {{Gilman}}},\ }in\ \href@noop {} {\emph {\bibinfo {booktitle} {Basic
  Mechanisms of Solar Activity}}},\ \bibinfo {series} {IAU Symposium},
  Vol.~\bibinfo {volume} {71},\ \bibinfo {editor} {edited by\ \bibinfo {editor}
  {\bibfnamefont {V.}~\bibnamefont {{Bumba}}}\ and\ \bibinfo {editor}
  {\bibfnamefont {J.}~\bibnamefont {{Kleczek}}}}\ (\bibinfo {year} {1976})\ p.\
  \bibinfo {pages} {207}\BibitemShut {NoStop}%
\bibitem [{\citenamefont {{Gilman}}(1977)}]{Gilman77}%
  \BibitemOpen
  \bibfield  {author} {\bibinfo {author} {\bibfnamefont {P.~A.}\ \bibnamefont
  {{Gilman}}},\ }\href@noop {} {\bibfield  {journal} {\bibinfo  {journal}
  {Geophysical and Astrophysical Fluid Dynamics}\ }\textbf {\bibinfo {volume}
  {8}},\ \bibinfo {pages} {93} (\bibinfo {year} {1977})}\BibitemShut {NoStop}%
\bibitem [{\citenamefont {{Gilman}}(1978{\natexlab{a}})}]{Gilman78}%
  \BibitemOpen
  \bibfield  {author} {\bibinfo {author} {\bibfnamefont {P.~A.}\ \bibnamefont
  {{Gilman}}},\ }\href@noop {} {\bibfield  {journal} {\bibinfo  {journal}
  {Geophysical and Astrophysical Fluid Dynamics}\ }\textbf {\bibinfo {volume}
  {11}},\ \bibinfo {pages} {157} (\bibinfo {year}
  {1978}{\natexlab{a}})}\BibitemShut {NoStop}%
\bibitem [{\citenamefont {{Christensen}}(2002)}]{Christensen}%
  \BibitemOpen
  \bibfield  {author} {\bibinfo {author} {\bibfnamefont {U.~R.}\ \bibnamefont
  {{Christensen}}},\ }\href@noop {} {\bibfield  {journal} {\bibinfo  {journal}
  {Journal of Fluid Mechanics}\ }\textbf {\bibinfo {volume} {470}},\ \bibinfo
  {pages} {115} (\bibinfo {year} {2002})}\BibitemShut {NoStop}%
\bibitem [{\citenamefont {{Heimpel}}\ \emph {et~al.}(2005)\citenamefont
  {{Heimpel}}, \citenamefont {{Aurnou}},\ and\ \citenamefont
  {{Wicht}}}]{Heimpel05}%
  \BibitemOpen
  \bibfield  {author} {\bibinfo {author} {\bibfnamefont {M.}~\bibnamefont
  {{Heimpel}}}, \bibinfo {author} {\bibfnamefont {J.}~\bibnamefont {{Aurnou}}},
  \ and\ \bibinfo {author} {\bibfnamefont {J.}~\bibnamefont {{Wicht}}},\
  }\href@noop {} {\bibfield  {journal} {\bibinfo  {journal} {\nat}\ }\textbf
  {\bibinfo {volume} {438}},\ \bibinfo {pages} {193} (\bibinfo {year}
  {2005})}\BibitemShut {NoStop}%
\bibitem [{\citenamefont {{Heimpel}}\ and\ \citenamefont
  {{Aurnou}}(2007)}]{HA07}%
  \BibitemOpen
  \bibfield  {author} {\bibinfo {author} {\bibfnamefont {M.}~\bibnamefont
  {{Heimpel}}}\ and\ \bibinfo {author} {\bibfnamefont {J.}~\bibnamefont
  {{Aurnou}}},\ }\href@noop {} {\bibfield  {journal} {\bibinfo  {journal}
  {Icarus}\ }\textbf {\bibinfo {volume} {187}},\ \bibinfo {pages} {540}
  (\bibinfo {year} {2007})}\BibitemShut {NoStop}%
\bibitem [{\citenamefont {{Aurnou}}\ \emph {et~al.}(2007)\citenamefont
  {{Aurnou}}, \citenamefont {{Heimpel}},\ and\ \citenamefont {{Wicht}}}]{AH07}%
  \BibitemOpen
  \bibfield  {author} {\bibinfo {author} {\bibfnamefont {J.}~\bibnamefont
  {{Aurnou}}}, \bibinfo {author} {\bibfnamefont {M.}~\bibnamefont {{Heimpel}}},
  \ and\ \bibinfo {author} {\bibfnamefont {J.}~\bibnamefont {{Wicht}}},\ }\href
  {\doibase 10.1016/j.icarus.2007.02.024} {\bibfield  {journal} {\bibinfo
  {journal} {Icarus}\ }\textbf {\bibinfo {volume} {190}},\ \bibinfo {pages}
  {110} (\bibinfo {year} {2007})}\BibitemShut {NoStop}%
\bibitem [{\citenamefont {{Cattaneo}}\ \emph {et~al.}(2003)\citenamefont
  {{Cattaneo}}, \citenamefont {{Emonet}},\ and\ \citenamefont
  {{Weiss}}}]{Cat03}%
  \BibitemOpen
  \bibfield  {author} {\bibinfo {author} {\bibfnamefont {F.}~\bibnamefont
  {{Cattaneo}}}, \bibinfo {author} {\bibfnamefont {T.}~\bibnamefont
  {{Emonet}}}, \ and\ \bibinfo {author} {\bibfnamefont {N.}~\bibnamefont
  {{Weiss}}},\ }\href@noop {} {\bibfield  {journal} {\bibinfo  {journal}
  {\apj}\ }\textbf {\bibinfo {volume} {588}},\ \bibinfo {pages} {1183}
  (\bibinfo {year} {2003})}\BibitemShut {NoStop}%
\bibitem [{\citenamefont {{Miesch}}(2001)}]{Miesch2001}%
  \BibitemOpen
  \bibfield  {author} {\bibinfo {author} {\bibfnamefont {M.~S.}\ \bibnamefont
  {{Miesch}}},\ }\href@noop {} {\bibfield  {journal} {\bibinfo  {journal}
  {\apj}\ }\textbf {\bibinfo {volume} {562}},\ \bibinfo {pages} {1058}
  (\bibinfo {year} {2001})}\BibitemShut {NoStop}%
\bibitem [{Note1()}]{Note1}%
  \BibitemOpen
  \bibinfo {note} {We numerically solve the non-dimensional Boussinesq
  equations in which we have used the outer radius as the lengthscale. If we
  wanted to compare spherical numerical simulations with Cartesian ones, we
  would have to non-dimensionalize the problem using the thickness of the shell
  $[l]=r_o-r_i=L$ such that both the problems could have the same effective
  Rayleigh number for accurate comparison.}\BibitemShut {Stop}%
\bibitem [{\citenamefont {{Aubert}}\ \emph {et~al.}(2008)\citenamefont
  {{Aubert}}, \citenamefont {{Aurnou}},\ and\ \citenamefont
  {{Wicht}}}]{parody}%
  \BibitemOpen
  \bibfield  {author} {\bibinfo {author} {\bibfnamefont {J.}~\bibnamefont
  {{Aubert}}}, \bibinfo {author} {\bibfnamefont {J.}~\bibnamefont {{Aurnou}}},
  \ and\ \bibinfo {author} {\bibfnamefont {J.}~\bibnamefont {{Wicht}}},\
  }\href@noop {} {\bibfield  {journal} {\bibinfo  {journal} {Geophysical
  Journal International}\ }\textbf {\bibinfo {volume} {172}},\ \bibinfo {pages}
  {945} (\bibinfo {year} {2008})}\BibitemShut {NoStop}%
\bibitem [{\citenamefont {{Christensen-Dalsgaard}}\ \emph
  {et~al.}(1996)\citenamefont {{Christensen-Dalsgaard}}, \citenamefont
  {{Dappen}}, \citenamefont {{Ajukov}}, \citenamefont {{Anderson}},
  \citenamefont {{Antia}}, \citenamefont {{Basu}}, \citenamefont {{Baturin}},
  \citenamefont {{Berthomieu}}, \citenamefont {{Chaboyer}}, \citenamefont
  {{Chitre}}, \citenamefont {{Cox}}, \citenamefont {{Demarque}}, \citenamefont
  {{Donatowicz}}, \citenamefont {{Dziembowski}}, \citenamefont {{Gabriel}},
  \citenamefont {{Gough}}, \citenamefont {{Guenther}}, \citenamefont {{Guzik}},
  \citenamefont {{Harvey}}, \citenamefont {{Hill}}, \citenamefont {{Houdek}},
  \citenamefont {{Iglesias}}, \citenamefont {{Kosovichev}}, \citenamefont
  {{Leibacher}}, \citenamefont {{Morel}}, \citenamefont {{Proffitt}},
  \citenamefont {{Provost}}, \citenamefont {{Reiter}}, \citenamefont
  {{Rhodes}}, \citenamefont {{Rogers}}, \citenamefont {{Roxburgh}},
  \citenamefont {{Thompson}},\ and\ \citenamefont {{Ulrich}}}]{ModelS}%
  \BibitemOpen
  \bibfield  {author} {\bibinfo {author} {\bibfnamefont {J.}~\bibnamefont
  {{Christensen-Dalsgaard}}}, \bibinfo {author} {\bibfnamefont
  {W.}~\bibnamefont {{Dappen}}}, \bibinfo {author} {\bibfnamefont {S.~V.}\
  \bibnamefont {{Ajukov}}}, \bibinfo {author} {\bibfnamefont {E.~R.}\
  \bibnamefont {{Anderson}}}, \bibinfo {author} {\bibfnamefont {H.~M.}\
  \bibnamefont {{Antia}}}, \bibinfo {author} {\bibfnamefont {S.}~\bibnamefont
  {{Basu}}}, \bibinfo {author} {\bibfnamefont {V.~A.}\ \bibnamefont
  {{Baturin}}}, \bibinfo {author} {\bibfnamefont {G.}~\bibnamefont
  {{Berthomieu}}}, \bibinfo {author} {\bibfnamefont {B.}~\bibnamefont
  {{Chaboyer}}}, \bibinfo {author} {\bibfnamefont {S.~M.}\ \bibnamefont
  {{Chitre}}}, \bibinfo {author} {\bibfnamefont {A.~N.}\ \bibnamefont {{Cox}}},
  \bibinfo {author} {\bibfnamefont {P.}~\bibnamefont {{Demarque}}}, \bibinfo
  {author} {\bibfnamefont {J.}~\bibnamefont {{Donatowicz}}}, \bibinfo {author}
  {\bibfnamefont {W.~A.}\ \bibnamefont {{Dziembowski}}}, \bibinfo {author}
  {\bibfnamefont {M.}~\bibnamefont {{Gabriel}}}, \bibinfo {author}
  {\bibfnamefont {D.~O.}\ \bibnamefont {{Gough}}}, \bibinfo {author}
  {\bibfnamefont {D.~B.}\ \bibnamefont {{Guenther}}}, \bibinfo {author}
  {\bibfnamefont {J.~A.}\ \bibnamefont {{Guzik}}}, \bibinfo {author}
  {\bibfnamefont {J.~W.}\ \bibnamefont {{Harvey}}}, \bibinfo {author}
  {\bibfnamefont {F.}~\bibnamefont {{Hill}}}, \bibinfo {author} {\bibfnamefont
  {G.}~\bibnamefont {{Houdek}}}, \bibinfo {author} {\bibfnamefont {C.~A.}\
  \bibnamefont {{Iglesias}}}, \bibinfo {author} {\bibfnamefont {A.~G.}\
  \bibnamefont {{Kosovichev}}}, \bibinfo {author} {\bibfnamefont {J.~W.}\
  \bibnamefont {{Leibacher}}}, \bibinfo {author} {\bibfnamefont
  {P.}~\bibnamefont {{Morel}}}, \bibinfo {author} {\bibfnamefont {C.~R.}\
  \bibnamefont {{Proffitt}}}, \bibinfo {author} {\bibfnamefont
  {J.}~\bibnamefont {{Provost}}}, \bibinfo {author} {\bibfnamefont
  {J.}~\bibnamefont {{Reiter}}}, \bibinfo {author} {\bibfnamefont {E.~J.}\
  \bibnamefont {{Rhodes}}, \bibfnamefont {Jr.}}, \bibinfo {author}
  {\bibfnamefont {F.~J.}\ \bibnamefont {{Rogers}}}, \bibinfo {author}
  {\bibfnamefont {I.~W.}\ \bibnamefont {{Roxburgh}}}, \bibinfo {author}
  {\bibfnamefont {M.~J.}\ \bibnamefont {{Thompson}}}, \ and\ \bibinfo {author}
  {\bibfnamefont {R.~K.}\ \bibnamefont {{Ulrich}}},\ }\href@noop {} {\bibfield
  {journal} {\bibinfo  {journal} {Science}\ }\textbf {\bibinfo {volume}
  {272}},\ \bibinfo {pages} {1286} (\bibinfo {year} {1996})}\BibitemShut
  {NoStop}%
\bibitem [{\citenamefont {{K{\"a}pyl{\"a}}}\ \emph {et~al.}(2017)\citenamefont
  {{K{\"a}pyl{\"a}}}, \citenamefont {{Rheinhardt}}, \citenamefont
  {{Brandenburg}}, \citenamefont {{Arlt}}, \citenamefont {{K{\"a}pyl{\"a}}},
  \citenamefont {{Lagg}}, \citenamefont {{Olspert}},\ and\ \citenamefont
  {{Warnecke}}}]{Kapyla}%
  \BibitemOpen
  \bibfield  {author} {\bibinfo {author} {\bibfnamefont {P.~J.}\ \bibnamefont
  {{K{\"a}pyl{\"a}}}}, \bibinfo {author} {\bibfnamefont {M.}~\bibnamefont
  {{Rheinhardt}}}, \bibinfo {author} {\bibfnamefont {A.}~\bibnamefont
  {{Brandenburg}}}, \bibinfo {author} {\bibfnamefont {R.}~\bibnamefont
  {{Arlt}}}, \bibinfo {author} {\bibfnamefont {M.~J.}\ \bibnamefont
  {{K{\"a}pyl{\"a}}}}, \bibinfo {author} {\bibfnamefont {A.}~\bibnamefont
  {{Lagg}}}, \bibinfo {author} {\bibfnamefont {N.}~\bibnamefont {{Olspert}}}, \
  and\ \bibinfo {author} {\bibfnamefont {J.}~\bibnamefont {{Warnecke}}},\
  }\href@noop {} {\bibfield  {journal} {\bibinfo  {journal} {ArXiv e-prints}\ }
  (\bibinfo {year} {2017})},\ \Eprint {http://arxiv.org/abs/1703.06845}
  {arXiv:1703.06845 [astro-ph.SR]} \BibitemShut {NoStop}%
\bibitem [{\citenamefont {{Otero}}\ \emph {et~al.}(2002)\citenamefont
  {{Otero}}, \citenamefont {{Wittenberg}}, \citenamefont {{Worthing}},\ and\
  \citenamefont {{Doering}}}]{Otero2002}%
  \BibitemOpen
  \bibfield  {author} {\bibinfo {author} {\bibfnamefont {J.}~\bibnamefont
  {{Otero}}}, \bibinfo {author} {\bibfnamefont {R.~W.}\ \bibnamefont
  {{Wittenberg}}}, \bibinfo {author} {\bibfnamefont {R.~A.}\ \bibnamefont
  {{Worthing}}}, \ and\ \bibinfo {author} {\bibfnamefont {C.~R.}\ \bibnamefont
  {{Doering}}},\ }\href@noop {} {\bibfield  {journal} {\bibinfo  {journal}
  {Journal of Fluid Mechanics}\ }\textbf {\bibinfo {volume} {473}},\ \bibinfo
  {pages} {191} (\bibinfo {year} {2002})}\BibitemShut {NoStop}%
\bibitem [{\citenamefont {{Johnston}}\ and\ \citenamefont
  {{Doering}}(2009)}]{Johnston}%
  \BibitemOpen
  \bibfield  {author} {\bibinfo {author} {\bibfnamefont {H.}~\bibnamefont
  {{Johnston}}}\ and\ \bibinfo {author} {\bibfnamefont {C.~R.}\ \bibnamefont
  {{Doering}}},\ }\href@noop {} {\bibfield  {journal} {\bibinfo  {journal}
  {Physical Review Letters}\ }\textbf {\bibinfo {volume} {102}},\ \bibinfo
  {eid} {064501} (\bibinfo {year} {2009})}\BibitemShut {NoStop}%
\bibitem [{\citenamefont {{Gilman}}(1978{\natexlab{b}})}]{G78temp}%
  \BibitemOpen
  \bibfield  {author} {\bibinfo {author} {\bibfnamefont {P.~A.}\ \bibnamefont
  {{Gilman}}},\ }\href@noop {} {\bibfield  {journal} {\bibinfo  {journal}
  {Geophysical and Astrophysical Fluid Dynamics}\ }\textbf {\bibinfo {volume}
  {11}},\ \bibinfo {pages} {157} (\bibinfo {year}
  {1978}{\natexlab{b}})}\BibitemShut {NoStop}%
\bibitem [{\citenamefont {{Verzicco}}\ and\ \citenamefont
  {{Sreenivasan}}(2008)}]{VZ08}%
  \BibitemOpen
  \bibfield  {author} {\bibinfo {author} {\bibfnamefont {R.}~\bibnamefont
  {{Verzicco}}}\ and\ \bibinfo {author} {\bibfnamefont {K.~R.}\ \bibnamefont
  {{Sreenivasan}}},\ }\href@noop {} {\bibfield  {journal} {\bibinfo  {journal}
  {Journal of Fluid Mechanics}\ }\textbf {\bibinfo {volume} {595}},\ \bibinfo
  {pages} {203} (\bibinfo {year} {2008})}\BibitemShut {NoStop}%
\bibitem [{\citenamefont {{Stevens}}\ \emph {et~al.}(2011)\citenamefont
  {{Stevens}}, \citenamefont {{Lohse}},\ and\ \citenamefont
  {{Verzicco}}}]{SLZ11}%
  \BibitemOpen
  \bibfield  {author} {\bibinfo {author} {\bibfnamefont {R.~J.~A.~M.}\
  \bibnamefont {{Stevens}}}, \bibinfo {author} {\bibfnamefont {D.}~\bibnamefont
  {{Lohse}}}, \ and\ \bibinfo {author} {\bibfnamefont {R.}~\bibnamefont
  {{Verzicco}}},\ }\href@noop {} {\bibfield  {journal} {\bibinfo  {journal}
  {Journal of Fluid Mechanics}\ }\textbf {\bibinfo {volume} {688}},\ \bibinfo
  {pages} {31} (\bibinfo {year} {2011})}\BibitemShut {NoStop}%
\bibitem [{\citenamefont {{Chan}}\ and\ \citenamefont
  {{Gigas}}(1992)}]{Chanetal}%
  \BibitemOpen
  \bibfield  {author} {\bibinfo {author} {\bibfnamefont {K.~L.}\ \bibnamefont
  {{Chan}}}\ and\ \bibinfo {author} {\bibfnamefont {D.}~\bibnamefont
  {{Gigas}}},\ }\href@noop {} {\bibfield  {journal} {\bibinfo  {journal}
  {Astrophys. J. Letters}\ }\textbf {\bibinfo {volume} {389}},\ \bibinfo
  {pages} {L87} (\bibinfo {year} {1992})}\BibitemShut {NoStop}%
\bibitem [{\citenamefont {{Moc{\'a}k}}\ \emph {et~al.}(2009)\citenamefont
  {{Moc{\'a}k}}, \citenamefont {{M{\"u}ller}}, \citenamefont {{Weiss}},\ and\
  \citenamefont {{Kifonidis}}}]{Mocak}%
  \BibitemOpen
  \bibfield  {author} {\bibinfo {author} {\bibfnamefont {M.}~\bibnamefont
  {{Moc{\'a}k}}}, \bibinfo {author} {\bibfnamefont {E.}~\bibnamefont
  {{M{\"u}ller}}}, \bibinfo {author} {\bibfnamefont {A.}~\bibnamefont
  {{Weiss}}}, \ and\ \bibinfo {author} {\bibfnamefont {K.}~\bibnamefont
  {{Kifonidis}}},\ }\href@noop {} {\bibfield  {journal} {\bibinfo  {journal}
  {Astronomy \& Astrophysics}\ }\textbf {\bibinfo {volume} {501}},\ \bibinfo
  {pages} {659} (\bibinfo {year} {2009})}\BibitemShut {NoStop}%
\bibitem [{\citenamefont {{Herwig}}\ \emph {et~al.}(2006)\citenamefont
  {{Herwig}}, \citenamefont {{Freytag}}, \citenamefont {{Hueckstaedt}},\ and\
  \citenamefont {{Timmes}}}]{Herwig}%
  \BibitemOpen
  \bibfield  {author} {\bibinfo {author} {\bibfnamefont {F.}~\bibnamefont
  {{Herwig}}}, \bibinfo {author} {\bibfnamefont {B.}~\bibnamefont {{Freytag}}},
  \bibinfo {author} {\bibfnamefont {R.~M.}\ \bibnamefont {{Hueckstaedt}}}, \
  and\ \bibinfo {author} {\bibfnamefont {F.~X.}\ \bibnamefont {{Timmes}}},\
  }\href@noop {} {\bibfield  {journal} {\bibinfo  {journal} {\apj}\ }\textbf
  {\bibinfo {volume} {642}},\ \bibinfo {pages} {1057} (\bibinfo {year}
  {2006})}\BibitemShut {NoStop}%
\bibitem [{\citenamefont {{Brummell}}\ \emph {et~al.}(2017)\citenamefont
  {{Brummell}}, \citenamefont {{Garaud}},\ and\ \citenamefont
  {{Korre}}}]{Brummelletal}%
  \BibitemOpen
  \bibfield  {author} {\bibinfo {author} {\bibfnamefont {N.}~\bibnamefont
  {{Brummell}}}, \bibinfo {author} {\bibfnamefont {P.}~\bibnamefont
  {{Garaud}}}, \ and\ \bibinfo {author} {\bibfnamefont {L.}~\bibnamefont
  {{Korre}}},\ }\href@noop {} {\bibfield  {journal} {\bibinfo  {journal} {in
  prep.}\ } (\bibinfo {year} {2017})}\BibitemShut {NoStop}%
\end{thebibliography}%

\providecommand{\noopsort}[1]{}\providecommand{\singleletter}[1]{#1}%

\end{document}